\documentclass[prb,preprint,showpacs]{revtex4}% Physical Review B

\usepackage{graphicx}
\usepackage{amsmath,amssymb}
\begin{document}
\title{Two mini-band model for self-sustained oscillations of the current through resonant tunneling semiconductor superlattices}
\author{M. \'Alvaro and L.L. Bonilla}
\affiliation {G. Mill\'an Institute, Fluid Dynamics, Nanoscience and Industrial
Mathematics, Universidad Carlos III de Madrid, 28911 Legan\'es, Spain}
\date{\today}
\begin{abstract}
A two miniband model for electron transport in semiconductor superlattices that includes scattering and interminiband tunnelling is proposed. The
model is formulated in terms of Wigner functions in a basis spanned by Pauli matrices, includes electron-electron scattering in the Hartree
approximation and modified Bhatnagar-Gross-Krook collision tems. For strong applied fields, balance equations for the electric field and the miniband
populations are derived using a Chapman-Enskog perturbation technique. These equations are then solved numerically for a dc voltage biased
superlattice. Results include self-sustained current oscillations due to repeated nucleation of electric field pulses at the injecting contact region and their motion towards the collector. Numerical reconstruction of the Wigner functions shows that the miniband with higher energy is empty during most of the oscillation period: it becomes populated only when the local electric field (corresponding to the passing pulse) is sufficiently large to trigger resonant tunneling. %strongly coupled superlattices and, in the limit of a weakly coupled superlattice, an approximate sequential resonant tunnelling mechanism and formation of static electric field domains. Unlike in the case of the existing ad hoc spatially discrete superlattice models, these results are derived directly from quantum kinetic equations.
\end{abstract}
\maketitle %\vspace{-1.2in}

\section{Introduction}
\label{sec:0}
Consider a n-doped semiconductor superlattice (SL) under a sufficiently large vertical voltage bias so that electron transport is due to resonant
tunneling between minibands. For small voltage values, electron transport chiefly involves the lowest miniband and there are many appropriate
kinetic theory descriptions: semiclassical Boltzmann-type equations \cite{kss72,ign87,ign91,sib95,bep03}, density matrix formulations \cite{bk97,fis98}, transport
equations for the nonequilibrium Green function (NGF) \cite{HJ08}, and Wigner-Poisson (WP) equations \cite{be05}. Semiclassical equations are
easier to handle and, in particular, can be used to describe space-charge instabilities such as self-sustained oscillations of the current (SSOC) in
dc voltage biased SLs due to the formation and dynamics of electric field domains \cite{BGr05}. SSOC can be found by deriving and solving a
drift-diffusion system from the semiclassical kinetic equation \cite{bep03}, or by a direct numerical solution of the latter \cite{cbc09}. Quantum
transport description based on NGFs are still limited to spatially homogeneous electric fields and therefore cannot be used to describe properly
space-charge phenomena \cite{HJ08}. WP equations can be used to derive nonlocal drift-diffusion systems exhibiting SSOC provided collision terms are
of Bhatnagar-Gross-Krook (BGK) type \cite{be05}.

In contrast to work in one-miniband SL, much less is known about first-principles space-charge transport involving resonant tunneling in SL \cite{BGr05}. Most of the work on resonant tunneling SL assume a large separation between time scales such that electron density and electric field can be assumed to be constant in each SL period and the tunneling current across barriers can be assumed to be stationary. Then expressions for the stationary current in an infinitely long SL under a constant electric field can be calculated by any quantum kinetic method and inserted in discrete balance equations \cite{BGr05}. The resulting models have been vastly useful to understand nonlinear electron transport in SL but they have not been derived from first principles. Recently, we have found a consistent perturbation method to derive nonlocal drift-diffusion systems (NDDS) from WP descriptions of two-miniband SLs with Rashba spin-orbit interaction \cite{bba08}. However, coupling between minibands in that work does not contemplate resonant tunneling between them for the underlying physical description of the SL is too simple.

Some time ago, Morandi and Modugno studied a variant of the standard k-p theory in which interband coupling terms depend on the applied electric field and used it to study wave function dynamics of a resonant tunneling diode \cite{MM}. For the same system, multiband Wigner function approaches have also been considered \cite{unl04,dem02,mor08,mor09}. Unlu et al \cite{unl04} use a nonequilibrium Green function formulation that includes scattering due to weak coupling to a phonon bath to derive equations for the multiband Wigner functions. A treatment of space-dependent but time-independent NGF and Wigner functions in MOSFET can be found in Ref. \onlinecite{jia10}. The other works focused their attention in coherent transport under an external field and near the semiclassical limit, thereby ignoring scattering \cite{dem02,mor08,mor09}. In this paper, we present a simplified model of a two-miniband SL using a field dependent coupling between minibands similar to that introduced for resonant tunneling diodes\cite{MM}. We consider the corresponding WP system with BGK collision terms that include collision broadening and decay between minibands due to scattering. Electron-electron scattering is treated in the Hartree approximation through the Poisson equation. We are interested in the {\em hyperbolic} limit in which electric field effects, including field-dependent inter-miniband transitions, are as strong as the BGK collision terms and dominate electron transport. By using the Chapman-Enskog perturbation method, we derive nonlocal balance equations for the electron population of the minibands and the electric field that inherit the nonlocality of the quantum Wigner equation. Numerical solutions of these nonlocal equations allow us to reconstruct the time-resolved Wigner matrix and they exhibit resonant tunneling between minibands and SSOC. During SSCO, we show that the miniband with higher-energy is practically empty except when the local electric field is sufficiently large to allow resonant tunneling from the miniband with lowest energy. Our calculations provide a first-principles description of SSCO in a resonant tunneling SL under dc voltage bias.

The rest of the paper is organized as follows. Section \ref{sec:1} contains the Hamiltonian we use as the basis of our kinetic theory. The governing WPBGK equations for the Wigner functions are introduced in Section \ref{sec:2}. The derivation of nonlocal balance equations by the Chapman-Enskog method is given in Section \ref{sec:3}. Section \ref{sec:4} presents numerical results obtained by solving the nonlocal balance equations with appropriate boundary conditions for the contact regions and dc voltage bias. In particular, these solutions include SSCO. Finally Section \ref{sec:5} contains our conclusions and the Appendix is devoted to technical matters.

\setcounter{equation}{0}
\section{Model Hamiltonian}
\label{sec:1}
Let us assume that the total Hamiltonian describing our system is
\begin{equation}
\mathbf{H}_{\rm total} = \mathbf{H} + \mathbf{H}_{\rm sc},\label{ham}
\end{equation} 
where $\mathbf{H}_{\rm sc}$ represents scattering and $\mathbf{H}(x,-i\partial/\partial x)$ is a $2\times 2$ Hamiltonian $\mathbf{H}$ corresponding to a SL with two minibands of widths $\Delta_{1}$ and $\Delta_{2}$, gap energy $2g$ and SL period $l$, 
\begin{eqnarray}
\mathbf{H}(x,k) &=& \left(
\begin{array}{cc}
-\frac{\Delta_{2}}{2}(1-\cos kl) - e W(x) + g & e F l \delta\\
 e F l \delta & \frac{\Delta_{1}}{2}(1-\cos kl) - e W(x) - g
\end{array}
\right), \label{1}\\
& \equiv& [h_{0}(k)- e W(x)]\boldsymbol{\sigma}_{0}+\vec{h}(k)\cdot
\vec{\boldsymbol{\sigma}} + eFl\delta\,\boldsymbol{\sigma_1} \nonumber ,
%&&\mathbf{H}(x,k) = [h_{0}(k)- e W(x)]\boldsymbol{\sigma}_{0}+\vec{h}(k)\cdot
%\vec{\boldsymbol{\sigma}} - i \frac{\partial}{\partial x}(-e W(x))\frac{(-i \hbar P)}{2 m^* g}%\boldsymbol{\sigma_1} ] \label{1}\\
%&& \quad\equiv\left(
%\begin{array}{cc}
%(\alpha+\gamma)(1-\cos kl) - e W(x) + g & - i\beta\sin kl + e F l \delta\\
% i\beta\sin kl + e F l \delta & (\alpha-\gamma)(1-\cos kl) - e W(x) - g
%\end{array} - i \frac{\partial}{\partial x}(-e W(x))\frac{(-i \hbar P)}{2 m^* g}
%\right) \nonumber .
\end{eqnarray}
Here we have considered tight-binding dispersion relations for the minibands and $-e<0$, $W$ and $-F=-\partial W/\partial x$ are the electron charge, the electric potential, and the electric field, respectively. The electric potential $W$ in $\mathbf{H}$ describes electron-electron interaction in a self-consistent Hartree approximation.

The matrix Hamiltonian $\mathbf{H}$ can be written as a linear combination of the Pauli matrices
\begin{eqnarray}
&&\boldsymbol{\sigma}_{0}= \left(\begin{array}{cc}
1 & 0 \\
0 & 1
\end{array}\right),
\,\boldsymbol{\sigma}_{1}= \left(\begin{array}{cc}
0 & 1 \\
1 & 0
\end{array}\right),\,
\boldsymbol{\sigma}_{2}= \left(\begin{array}{cc}
0 & -i \\
i & 0
\end{array}\right),\,
\boldsymbol{\sigma}_{3}= \left(\begin{array}{cc}
1 & 0 \\
0 & -1
\end{array}\right), \label{3} \nonumber
\end{eqnarray}
with coefficients:
\begin{eqnarray}
\begin{array}{cc}
h_{0}(k)= -\alpha\, (1-\cos kl), & h_{1}(k)=0, \\
h_{2}(k)=0, & h_{3}(k)= -\gamma\, (1-\cos kl) + g,\\
\alpha=\frac{\Delta_{2}-\Delta_{1}}{4}, & \gamma=\frac{\Delta_{2}+
\Delta_{1}}{4}.
\end{array}  \label{2}
\end{eqnarray}
The term $eFl\delta\,\boldsymbol{\sigma}_1$ in (\ref{1}) is a field-dependent tunneling term derived by means of the k-p theory for the evolution of the Wannier envelope functions [cf.\ Equations (33) of Ref. \onlinecite{MM} without second order terms, i.e.\ with $M_{n n'}=0$]. The dimensionless parameter $\delta$ is a phenomenological parameter proportional to the interminiband momentum matrix element:
\begin{eqnarray}
\delta=\frac{\hbar P}{2 m^* g l}, \quad P = \frac{\hbar}{l}\int_{-l/2}^{l/2}
{u_2^* \frac{\partial{u_1}}{\partial{ x}}  dx}, \label{2bis}
\end{eqnarray}
where $u_{1,2}$ are the periodic parts of the miniband Bloch functions. A related model has been used to describe coherent transport in a resonant
interband tunnelling diode \cite{MM,mor08,mor09}.

The miniband energies ${\cal E}^{\pm}(k)$ are the eigenvalues of the free Hamiltonian $\mathbf{H}_{0}(k)=h_{0}(k)\boldsymbol{\sigma}_{0}+\vec{h}(k)\cdot
\vec{\boldsymbol{\sigma}}$ (zero electric potential), given by
\begin{equation}
\mathcal{E}^{\pm}(k) = h_{0}(k) \pm h_{3}(k). \label{4}
\end{equation}
The corresponding spectral projections are
\begin{equation}
\mathbf{P}^{\pm} = \frac{\boldsymbol{\sigma}_0\pm\boldsymbol{\sigma}_3}{2},
\label{5}
\end{equation}
so that we can write
\begin{equation}
\mathbf{H}_{0}(k)= \mathcal{E}^{+}(k)\mathbf{P}^{+} + \mathcal{E}^{-}(k)
\mathbf{P}^{-}.   \label{6}
\end{equation}

\section{Wigner function description}
\label{sec:2}
If $\psi_a(x,y,z,t)$, $a=1,2$, are the second quantized wave function amplitudes expressed in the Bloch basis, the Wigner matrix is \cite{bba08}
\begin{eqnarray}
f_{ab}(x,k,t)= {2l\over S}\sum_{j=-\infty}^\infty \int_{\mathbb{R}^2}\langle
\psi^\dagger_a(x+jl/2,y,z,t)\psi_b(x-jl/2,y,z,t)\rangle e^{ijkl} d\mathbf{x}_{\perp}  ,
\label{wigner}
\end{eqnarray}
where $S$ is the SL cross section. Note that the Wigner matrix is periodic in $k$ with period $2\pi/l$.
It is convenient to write the Wigner matrix $\mathbf{f}(x,k,t)$ in terms of the Pauli matrices:
\begin{equation}
\mathbf{f}(x,k,t) = \sum_{i=0}^{3} f^{i}(x,k,t)\boldsymbol{\sigma}_{i} =
f^0(x,k,t)\boldsymbol{\sigma}_{0} + \vec{f}(x,k,t)\cdot\vec{ \boldsymbol{\sigma}}. \label{7}
\end{equation}
The Wigner components $f^i(x,k,t)$ are real and can be related to the coefficients of the Hermitian
Wigner matrix by
\begin{eqnarray}
\begin{array}{cc}
f_{11}= f^0 + f^3, & f_{12}= f^1 - if^2, \\
f_{21}=f^1 + i f^2, & f_{22}= f^0 - f^3.
\end{array}  \label{8}
\end{eqnarray}
Hereinafter we shall use the equivalent notations
\begin{eqnarray}
f= \left(\begin{array}{c}
f^0 \\
\vec{f}
\end{array}\right) =
\left(\begin{array}{c}
f^0 \\
f^1\\
f^2\\
f^3
\end{array}\right). \label{9}
\end{eqnarray}
The populations of the minibands with energies $\mathcal{E}^\pm$
are given by the moments:
\begin{equation}
n^\pm(x,t) = {l\over 2\pi}\int_{-\pi/l}^{\pi/l} \left[ f^0(x,k,t)\pm f^3(x,k,t)\right]
\, dk, \label{10}
\end{equation}
and the total electron density is $n^+ + n^-$. 

After some algebra, from the time-dependent Schr\"odinger equations for wave functions $\psi_a$ with the Hamiltonian $\mathbf{H}_{\rm tot}$ in (\ref{ham}), we can obtain the following Wigner-Poisson-Bhatnagar-Gross-Krook (WPBGK) equations for the Wigner components
\begin{eqnarray}
&&{\partial f^0\over\partial t} - {\alpha\over\hbar}\sin kl\,\Delta^- f^0 -
{\gamma\over\hbar}\sin kl\,\Delta^- f^3 - \Theta_1 f^0 - \Theta_2 f^1 = Q^0[f],
\label{11}\\
&&{\partial\vec{f}\over\partial t} - {\alpha\over\hbar}\sin kl\,\Delta^-
\vec{f} - {\gamma\over\hbar}\sin kl\,\Delta^- f^0\,\vec{\nu} +
\vec{\omega}\times\vec{f} - \vec{\Theta}[f] = \vec{Q}[f], \label{12}
\end{eqnarray}
whose right hand sides contain collision terms $Q[f]$ arising from $\mathbf{H}_{\rm sc}$. These terms will be modeled phenomenologically and described later. Electron-electron collisions are treated in the Hartree approximation and described by the Poisson equation for the electrostatic potential:
\begin{eqnarray}
\varepsilon\, {\partial^2 W\over\partial x^2} = {e\over l}\, (n^+ + n^- -N_{D}),  \label{13}
\end{eqnarray} 
where $\varepsilon$ and $N_D$ are the SL permittivity and the  2D doping density, respectively. In (\ref{11}) - (\ref{12}),
\begin{eqnarray}
&&\vec{\omega} = {2(g-\gamma)+\gamma\cos kl\,\Delta^+\over\hbar}\,\vec{\nu},
\quad\vec{\nu}=  (0,0,1), \label{15}\\
&&\Theta_1 f^{m} (x,k,t)= {el\over i\hbar} \sum_{j=-\infty}^\infty j\langle F(x,t) \rangle_{j} e^{ijkl} f^{m}_{j}(x,t),\label{19}\\
&&\Theta_2 f^{m} (x,k,t)= -\frac{el \delta}{ i\hbar} \sum_{j=-\infty}^\infty e^{ijkl} f^{m}_{j}(x,t)\,\Delta^-_j F(x,t),\label{20}\\
&&\Theta_3 f^{m} (x,k,t)= \frac{el \delta}{ i\hbar}\sum_{j=-\infty}^\infty e^{ijkl} f^{m}_{j}(x,t)\,\Delta^+_j F(x,t),\label{21}\\
&&\vec{\Theta}[f]= \Theta_1\, \vec{f} + \left(%
\begin{array}{r}
  \Theta_2\, f^0 \\
  \Theta_3\, f^3 \\
  -\Theta_3\, f^2 \\
\end{array}
\right) .      \label{22theta}
\end{eqnarray}
We have defined the operators 
\begin{eqnarray}
(\Delta^\pm_j u)(x,k) = u\left(x+\frac{jl}{2},k\right)\pm u\left(x-\frac{jl}{2},k\right)   \label{14}
\end{eqnarray}
(the subscript is omitted for $j=1$) and the spatial averages:
\begin{eqnarray}
\langle F(x,t)\rangle_{j} &\equiv& {1\over jl} \int_{-jl/2}^{jl/2} F(x+s,t)\, ds \label{23average}\\
&=& \left\langle \frac{\partial W}{\partial x}(x,t)\right\rangle_j =\frac{\partial}{\partial x}\left\langle W(x,t)\right\rangle_j = \frac{\Delta^-_j W(x,t)}{jl}. \label{17}
\end{eqnarray}
Our collision model is similar to that used in Ref. \onlinecite{bba08} and it contains two terms: a BGK term which tries to send the miniband Wigner function to its local equilibrium and a scattering term that sends electrons from the miniband with higher energy (whose electron density is $n^+$) to the miniband with lower energy (whose electron density is $n^-$):
\begin{eqnarray}
&& Q^0[f] = - {f^0 - \Omega^0\over\tau},  \label{20q0}\\
&&\vec{Q}[f] = - {\vec{f} - \vec{\Omega}\over\tau} -
{\vec{\nu} f^0 + \vec{f}\over\tau_{\rm sc}}, \label{21qvec}\\
&&\Omega^{0} = {\phi^+ + \phi^-\over 2}\,, \quad
\vec{\Omega}  = {\phi^+ - \phi^-\over 2}\,\vec{\nu},\label{22}\\
&&\phi^{\pm}(k;n^\pm) = {m^{*}k_{B}T\over
\pi\hbar^2}\,\int_{-\infty}^\infty
\frac{\sqrt{2}\,\Gamma^3/\pi}{\Gamma^4 +[E-{\cal E}^\pm(k)]^4}\,
\ln\left(1+ e^{{\mu^\pm - E\over k_{B}T}}\right)\, dE,\label{23}\\
&& n^\pm = {l\over 2\pi}\int_{-\pi/l}^{\pi/l} \phi^\pm(k;n^\pm)\, dk. \label{24}
\end{eqnarray}
The chemical potentials of the minibands, $\mu^+$ and $\mu^{-}$ are calculated in terms of $n^+$ and $n^-$
respectively, by inserting (\ref{23}) in (\ref{24}) and solving the resulting equations. The local
equilibria $\phi^\pm$ are the integrals of collision-broadened 3D Fermi-Dirac distributions over
the lateral components of the wave vector on the plane perpendicular to the growth direction $x$.
\cite{bba08} As the broadening energy $\Gamma\to 0$, the line-width function in the integrand of
(\ref{23}) becomes $\delta(E-\mathcal{E}^\pm(k))$.

Our collision model should enforce charge continuity.
To check this, we first calculate the time derivative of $n^\pm$ using (\ref{10}) to (\ref{12}):
\begin{eqnarray}
{\partial n^\pm\over\partial t} - {\alpha l\Delta^-\over
2\pi\hbar} \int_{-\pi/l}^{\pi/l} \sin kl\,(f^0\pm f^3)\, dk -{\gamma l\Delta^-\over
2\pi\hbar}\int_{-\pi/l}^{\pi/l}\sin kl (f^3 \pm f^0)\, dk \nonumber\\
 \pm {l\over 2\pi}\int_{-\pi/l}^{\pi/l}\Theta_3 f^2\, dk
 = {l\over 2\pi}\int_{-\pi/l}^{\pi/l} (Q^0[f]\pm Q^3[f])\, dk = \mp {n^+
\over\tau_{\rm sc}},  \label{25}
\end{eqnarray}
where we have employed $\int \Theta_1 f^0 dk = \int \Theta_2 f^1 dk = 0$. Then we obtain:
\begin{eqnarray}
{\partial\over\partial t}(n^+ + n^-) - \Delta^-\left[ {l\over\pi\hbar}\int_{-\pi/
l}^{\pi/l} \sin kl\left(\alpha f^0 + \gamma f^3\right) dk \right] = 0. \label{26}
\end{eqnarray}
Noting that $\Delta^- u(x)= l\,\partial\langle u(x)\rangle_{1}/\partial x$, we see that
(\ref{26}) {\em is the charge continuity equation}. Differentiating in time the Poisson equation (\ref{13}),
using (\ref{26}) in the result and integrating with respect to $x$, we get the following nonlocal
Amp\`ere's law for the balance of current:
\begin{eqnarray}
\varepsilon {\partial F\over\partial t} - \left\langle {el\over\pi\hbar}
\int_{-\pi/l}^{\pi/l} \sin kl \left(\alpha f^0 + \gamma f^3\right) dk \right
\rangle_{1} = J(t).   \label{27}
\end{eqnarray}
Here the space independent function $J(t)$ is the total current density. Since the Wigner
components are real, we can rewrite (\ref{27}) in the following equivalent form:
\begin{eqnarray}
\varepsilon {\partial F\over\partial t} + {2e\over\hbar}\,\left\langle \alpha\,
\mbox{Im}f^0_{1} +\gamma\, \mbox{Im}f^3_{1}\right\rangle_{1} = J(t). \label{28}
\end{eqnarray}
We are using the notation $f_j^m$ for the Fourier coefficients of $f^m$:
\begin{eqnarray}
f^m(x,k,t)=\sum_{j=-\infty}^\infty f^m_{j}(x,t)\, e^{ijkl}. \label{fourier}
\end{eqnarray}

\section{The Chapman-Enskog method and balance equations}
\label{sec:3} 
In this Section, we shall derive the reduced balance equations for our two-miniband SL using the Chapman-Enskog method. Note that if we were to know the Wigner matrix as a function of $n^\pm$ and the electric field, Equations (\ref{25}) and the Poisson equation (\ref{13}) would be the sought balance equations and could be solved directly. As they are now, Equations (\ref{25}) are not closed. However, in a limit in which collisions and electric potential terms dominate all others in the Wigner equations, it is possible to use perturbation theory to close (\ref{25}). The idea is that in this so-called {\em hyperbolic limit}, the Wigner matrix is very close to a local equilibrium (modified by the electric field) which depends on $n^\pm$ and $F$. Using two terms in a Chapman-Enskog expansion, we show below that Equations (\ref{25}) can be closed.

First of all,
we should decide the order of magnitude of the terms in the WPBGK equations (\ref{11}) and (\ref{12}) in the hyperbolic limit. In this limit, the
collision frequency $1/\tau$ and the Bloch frequency $eF_{M}l/\hbar$ are of the same order, say about 10 THz. Then $F_{M}=O(\hbar/(el\tau))$.
Typically, $2g/\hbar$ is of the same order, so that the term containing $2g/\hbar$ in (\ref{12}) should also balance the BGK collision term. The
other terms are of order $\gamma l/(\hbar x_{0})$, where $x_{0}$ is the characteristic length over which the field varies, and they are much smaller,
so that $\lambda=\gamma\tau l/(\hbar x_{0})\ll 1$. From the Poisson equation, we obtain $x_{0}/l=\varepsilon F_{M}/(eN_{D})=\varepsilon \hbar/(e^2
\tau l N_{D})$, and therefore the small dimensionless parameter is
\begin{equation}
\lambda = \frac{e^2\tau^2\gamma l N_{D}}{\varepsilon\hbar^2}. \label{16}
\end{equation}
The scattering time $\tau_{\rm sc}$ is  much longer than the
collision time $\tau$, and we shall consider $\tau/\tau_{\rm sc}=
O(\lambda)\ll 1$.
Equations (\ref{11}) and (\ref{12}) can be written as the scaled WPBGK equations as follows:
\begin{eqnarray}
\mathbb{L} f -\Omega = - \lambda\,\left(\tau\,{\partial f\over\partial t}+ \Lambda f\right).
\label{30}
\end{eqnarray}
where we have inserted the book-keeping parameter $\lambda$ which is set equal to 1 at
the end of our calculations. \cite{bep03,bba08} This trick saves us from rewriting our equations in
nondimensional units. Here the operators $\mathbb{L}$ and $\Lambda$ are defined by
\begin{eqnarray}
&&\mathbb{L} f= f - \tau\, \Theta_1 f - \tau\, \Theta_2\left(%
\begin{array}{c}
  f^1 \\
  f^0 \\
  0 \\
  0 \\
\end{array}%
\right) - \tau\, \Theta_3\left(%
\begin{array}{c}
  0 \\
  0 \\
  f^3 \\
  -f^2 \\
\end{array}%
\right)   + \eta_{1} \left(
\begin{array}{c}
0\\
- f^2\\
f^{1}\\
0
\end{array}\right),  \label{31}\\
&& \Lambda f = \eta_{2}\, \left( \begin{array}{c}
0\\
\vec{f} + \vec{\nu} f^0
\end{array}\right) - {\tau\over\hbar}\sin kl\,\Delta^- \left[\alpha f
+ \gamma \left( \begin{array}{c}
f^3\\
\vec{\nu} f^0 \end{array}\right)\right] + \frac{\gamma\tau}{\hbar}
(\cos kl\, \Delta^+ -2)\left(
\begin{array}{c}
0\\
\vec{\nu}\times\vec{f} \end{array}\right),  \nonumber
\end{eqnarray}
where
\begin{equation}
\eta_{1}= {2g\tau\over\hbar},\quad \eta_{2}= {\tau\over\tau_{\rm sc}}. \label{33}
\end{equation}

To derive the reduced balance equations, we use the following
Chapman-Enskog ansatz:
\begin{eqnarray}
&& f(x,k,t;\epsilon) = f^{(0)}(k;n^+,n^-,F) + \sum_{m=1}^{\infty}
f^{(m)}(k;n^+,n^-,F)\, \lambda^{m} ,    \label{34}\\
&& \varepsilon {\partial F\over\partial t} + \sum_{m=0}^{\infty}
J_{m}(n^+,n^-,F)\, \lambda^{m} = J(t),  \label{35}\\
&&{\partial n^\pm\over\partial t} = \sum_{m=0}^{\infty} A^{\pm}_{m}(n^+,n^-,F)\, \lambda^{m}.
\label{36}
\end{eqnarray}
The functions $A_{m}^\pm$ and $J_{m}$ are related through the Poisson equation (\ref{13}), so that
\begin{eqnarray}
A^{+}_{m}+ A^-_{m} = - {l\over e} \, {\partial J_{m}\over\partial x}. \label{37}
\end{eqnarray}
Inserting (\ref{34}) to (\ref{36}) into (\ref{30}), we get
\begin{eqnarray}
&& \mathbb{L} f^{(0)} = \Omega, \label{38}\\
&& \mathbb{L} f^{(1)} = - \left.\tau\, {\partial f^{(0)}\over\partial t}\right|_{0} - \Lambda f^{(0)},\label{39}\\
&&\mathbb{L} f^{(2)} = - \left.\tau\, {\partial f^{(1)}\over\partial t}\right|_{0} -
\Lambda f^{(1)} - \left.\tau\,{\partial f^{(0)}\over\partial t}\right|_{1}, \label{40}
\end{eqnarray}
and so on. The subscripts 0 and 1 in the right hand side of these equations mean that we replace $\varepsilon\,\partial
F/\partial t|_{m}= J \delta_{0m}-J_{m}$, $\partial n^\pm/\partial t|_{m}=A^\pm_{m}$, provided $\delta_{00}=1$ and
$\delta_{0m}=0$ if $m \neq 0$. Moreover, inserting (\ref{34}) into (\ref{10}) yields the following
compatibility conditions:
\begin{eqnarray}
&& f^{(1)\,0}_{0}= f^{(1)\, 3}_{0} = 0,\label{41}\\
&& f^{(2)\, 0}_{0} = f^{(2)\, 3}_{0} = 0,\label{42}
\end{eqnarray}
etc.

To solve (\ref{38}) for $f^{(0)}\equiv\varphi$, we first note that
\begin{eqnarray}
&&-\tau\, \Theta_1\varphi =  i\sum_{j=-\infty}^\infty \vartheta_{j} \varphi_{j} e^{ijkl},  \label{43}\\
&&-\tau\, \Theta_2\varphi = - i \delta \,\sum_{j=-\infty}^\infty \varphi_{j} e^{ijkl}\Delta^-_j \mathcal F ,  \label{44}\\
&&-\tau\, \Theta_3\varphi = - \delta \,\sum_{j=-\infty}^\infty \varphi_{j} e^{ijkl}\Delta^+_j \mathcal F ,  \label{45}\\
&&\mathcal F \equiv {\tau el\over\hbar}\, F,\quad \vartheta_{j} \equiv j\,\langle {\mathcal F}\rangle_{j}.\label{46}
\end{eqnarray}
Then (\ref{38}) and (\ref{22}) yield
\begin{eqnarray}
&&\varphi_{j}^0 =  {\phi^+_{j} + \phi^-_{j}\over 2}\, \left[{1 \over 1+i \vartheta_j} -  \eta_1 \delta^2 Z_j\, M_j^+
(\Delta_j^-\mathcal F)^2 \right] \label{47} \\
&&\quad\quad +\, i{\phi_j^+-\phi_j^- \over 2} \eta_1 \delta^2Z_j\, (\Delta_j^- \mathcal F)\, (\Delta_j^+ \mathcal F) , \nonumber\\
&&\varphi_{j}^1 = {1\over2}\eta_1\delta(1+i\vartheta_j)\, Z_j \, \left[(\phi^+_{j} + \phi^-_{j})\, i M_j^+
 \Delta_j^-\mathcal F + (\phi^+_{j} - \phi^-_{j})\, \Delta_j^+ \mathcal F \right], \label{48}\\
 &&\varphi_{j}^2 = -{1\over2}\eta_1\delta(1+i\vartheta_j)\, Z_j \, \left[(\phi^+_{j} + \phi^-_{j})\, i
 \Delta_j^-\mathcal F - (\phi^+_{j} - \phi^-_{j})\, M_j^-\, \Delta_j^+ \mathcal F \right], \label{49}\\
&&\varphi_{j}^3 =  {\phi^+_{j} - \phi^-_{j}\over 2}\, \left[{1 \over 1+i \vartheta_j} -  \eta_1 \delta^2 Z_j\,M_j^-
(\Delta_j^+\mathcal F)^2 \right] \label{50}\\
&&\quad\quad +\, i{\phi_j^+ + \phi_j^- \over 2} \eta_1 \delta^2 Z_j\, (\Delta_j^- \mathcal F)\, (\Delta_j^+ \mathcal
F) . \nonumber
\end{eqnarray}
Here we have used that the Fourier coefficients
\begin{eqnarray}
\phi_{j}^\pm =  {l\over \pi}\, \int_{0}^{\pi/l}\cos(jkl)\,\phi^\pm\, dk, \label{51}
\end{eqnarray}
are real because $\phi^\pm$ are even functions of $k$. The coefficients $Z_j$ and $M_j^{\pm}$ are defined as
\begin{eqnarray}
&&M_j^{\pm} \equiv {1\over \eta_1}\left[1 + i \vartheta_j + {\delta^2(\Delta_j^{\pm}\mathcal F)^2 \over 1 +
i\vartheta_j} \right], \label{52} \\
&&Z_j \equiv {1\over \eta_1^2\,(1+i\vartheta_j)^2\, (1+M_j^+\,M_j^-)}. \label{53}
\end{eqnarray}
The solution $f^{(0)}=\varphi$ given by (\ref{47})-(\ref{50}) is essentially the
local equilibrium $\Omega$ given by (\ref{22})-(\ref{24}) modified by the field-dependent terms $\Theta_i$ that appear in the Wigner equations (\ref{11}) and (\ref{12}). This solution yields convective terms in the balance equations which contain first order differences. In the semiclassical limit, these equations become a hyperbolic system which may have discontinuous solutions (shock waves). Then it is convenient to regularize such solutions by keeping diffusion-like terms (second order differences) arising from the next-order Wigner functions $f^{(1)}$. 

The solution of (\ref{39}) is $f^{(1)}\equiv \psi$ with
\begin{eqnarray}
&& \psi_{j}^0 = {r_j^0 \over 1+i\vartheta_j}\left[1 - {\delta^2 M_j^+(\Delta_j^- \mathcal F)^2 \over \eta_1
(1+i\vartheta_j)(1+M_j^+M_j^-)}\right] \label{54}\\
&&\quad + {i\delta \Delta_j^- \mathcal F \over \eta_1(1+i\vartheta_j)(1+M_j^+M_j^-)}\left[M_j^+ r_j^1 + r_j^2 + {
\delta \Delta_j^+ \mathcal F \over 1 + i \vartheta_j}\, r_j^3 \right], \nonumber \\
&& \psi_{j}^1 = {1 \over \eta_1(1+M_j^+ M_j^-)}\left[M_j^+r_j^1 + {i\delta M_j^+\,\Delta_j^- \mathcal F \over 1 +
i \vartheta_j}\, r_j^0 + r_j^2 + {\delta\, \Delta_j^+ \mathcal F \over 1+i\vartheta_j }\,r_j^3 \right], \label{55}\\
&& \psi_{j}^2 = {1 \over \eta_1(1+M_j^+M_j^-)}\left[M_j^-\,r_j^2 + {\delta\,M_j^-\,\Delta_j^+ \mathcal F \over 1+i
\vartheta_j}\,r_j^3 - r_j^1 - {i\delta\, \Delta_j^- \mathcal F \over 1+i\vartheta_j}\,r_j^0 \right], \label{56}\\
&& \psi_{j}^3 = {r_j^3 \over 1+i\vartheta_j}\left[1 - {\delta^2 M_j^-(\Delta_j^+ \mathcal F)^2 \over \eta_1
(1+i\vartheta_j)(1+M_j^+M_j^-)}\right] \label{57}\\
&&\quad - {\delta \Delta_j^+ \mathcal F \over \eta_1(1+i\vartheta_j)(1+M_j^+M_j^-)}\left[M_j^-\,r_j^2 - r_j^1 - {i
\delta \Delta_j^- \mathcal F \over 1 + i \vartheta_j}\, r_j^0 \right]. \nonumber
\end{eqnarray}
Here $r$ is the right hand side of (\ref{39}). 

The balance equations can be found in two ways. We can calculate $A^\pm_m$ for $m = 0,1$ in (\ref{36}) by using the solvability conditions (\ref{41}) and (\ref{42}) in (\ref{39}) and (\ref{40}), respectively. More simply, we can obtain the balance equations by inserting the solutions (\ref{47}) to (\ref{50}) and (\ref{54}) to (\ref{57}) in the balance equations (\ref{25}) and in the Amp\`ere's law (\ref{27}). The result is:
\begin{eqnarray}
&& {\partial n^\pm\over\partial t} + \Delta^- D_{\pm}(n^+,n^-,F)= \mp R(n^+,n^-,F),\label{58}\\
&& \varepsilon\,{\partial F\over\partial t}+ e\,\left\langle D_+(n^+,n^-,F) + D_-(n^+,n^-,F) \right\rangle_{1} = J(t)\label{59}\\
&&D_{\pm} = {\alpha\pm\gamma\over\hbar}\,\mbox{Im}(\varphi_1^0 \pm \varphi_1^3 + \psi_1^0 \pm \psi_1^3),\label{60}\\
&& R = {1 \over \tau}\left[\eta_{2}n^+ + 2\delta \mathcal F (\varphi_0^2 + \psi_0^2)\right] .\label{61}
\end{eqnarray}
Note that Eq.\ (\ref{59}) can be obtained from (\ref{58}) and the Poisson equation. Equations (\ref{58}) to (\ref{61})
must be solved together with the Poisson equation (\ref{13}), the expression for the local equilibrium Wigner densities (\ref{23}) and expressions (\ref{24}) for $n^\pm$. The zeroth and first order Wigner functions $\varphi_j$ and $\psi_j$ in
(\ref{60}) and (\ref{61}) can be obtained from Equations (\ref{47})-(\ref{50}) and (\ref{54})- \ref{57}), respectively. The complete expressions for $D_{\pm}$ and $R$ are shown in Appendix \ref{appA}.

\section{Numerical results}
\label{sec:4}
To solve numerically the system of equations (\ref{58}) - (\ref{61}), we have to add the voltage bias conditions for the electric potential and appropriate boundary conditions at the contact regions. Note that our equations involve finite differences and several one-period integral averages. This means that we need to give boundary conditions over intervals of size $2l$ before $x=0$ and after $x=Nl$, not just boundary conditions at $x=0, Nl$ as we would give for semiclassical drift-diffusion equations. At the injecting region (cathode), the usual boundary condition is that the electron current density satisfies Ohm's law and therefore it is proportional to the electric field there. We use this condition for each point of the interval $-2l\leq x\leq 0$. Similarly, we also need the electron densities $n^\pm$ at the cathode. To avoid inconvenient boundary layer effects, we choose their values for a spatially uniform stationary state with a given value of the field. The resulting
  boundary conditions in $-2l\leq x \leq 0$ are: $W=0$ and
\begin{eqnarray}
&& \varepsilon\,{\partial F\over\partial t} + \sigma_{cathode}\,F = J,  \label{62}\\
&& n^\pm = n^{\pm}_{st},  \label{63}
\end{eqnarray}
where $n^{\pm}_{st}$ are the miniband electron densities corresponding to a spatially uniform stationary state. The latter can be obtained by equating to zero the right hand sides of the rate equation (\ref{58}) and the Poisson equation (\ref{13}): $R(n^+,n^-,F)=0$ and $n^++n^-=N_D$, respectively. The result is
\begin{eqnarray}
&& n^{\pm}_{st} = N_D\left({1\over2} \mp {\eta_2(1 + \eta_1^2 + 4\delta^2\mathcal F^2) \over 8\delta^2\mathcal F^2 + 2\eta_2(1 + \eta_1^2 +
4\delta^2\mathcal F^2)}\right).
\end{eqnarray}

The  boundary conditions in the anode region ($Nl\leq x \leq Nl+2l$) are: $W=V$ and
\begin{eqnarray}
&& \varepsilon\,{\partial F\over\partial t} + \sigma_{anode}\,({n^++n^- \over N_D})\,F = J,  \label{62anode}\\
&& n^+ = 0.
\end{eqnarray}
The lower miniband electron density $n^-$ in the anode region is obtained from the Poisson equation (\ref{13}).

To present numerical results, we have used the parameter values corresponding to a GaAs/AlAs SL from Table I of \onlinecite{khp97} which has narrow minibands so that resonant tunneling plays an important role in electron transport. Our parameter values are:  $d_{B}=1.5$ nm, $d_{W}=9$ nm, $l= d_{B}+d_{W} = 10.5$ nm, $N_{D}= 2.5\times 10^{10}$ cm$^{-2}$, $\tau= 0.0556$ ps, $\tau_{\rm
sc}= 0.556$ ps,\cite{shah} $V=9$ V, $N=200$, $\sigma_{cathode}= 1.4\, \Omega^{-1}$m$^{-1}$, $\sigma_{anode}= 0.7\, \Omega^{-1}$m$^{-1}$, $T=5$ K, $\Delta_1= 2.6$ meV, $\Delta_2= 13.2$ meV, $P/\hbar=0.2238$/nm,\cite{P} $\Gamma = 1$ meV.\cite{bba08} With these values, $\alpha = 2.6$ meV, $\gamma = 3.9$ meV, $\delta=0.12$. We have selected the following units to present our results graphically: $F_M=\hbar/(el\tau)= 11.28$ kV/cm, $x_{0} = \varepsilon F_M l /(eN_D) =31.4$ nm, $t_{0} = \hbar/\alpha =  0.25$ ps, $J_{0} = \alpha e N_D/\hbar = 1.58\times 10^4$ A/cm$^2$.\\
\begin{figure}
\begin{center}
\includegraphics[width=5.5cm,angle=0]{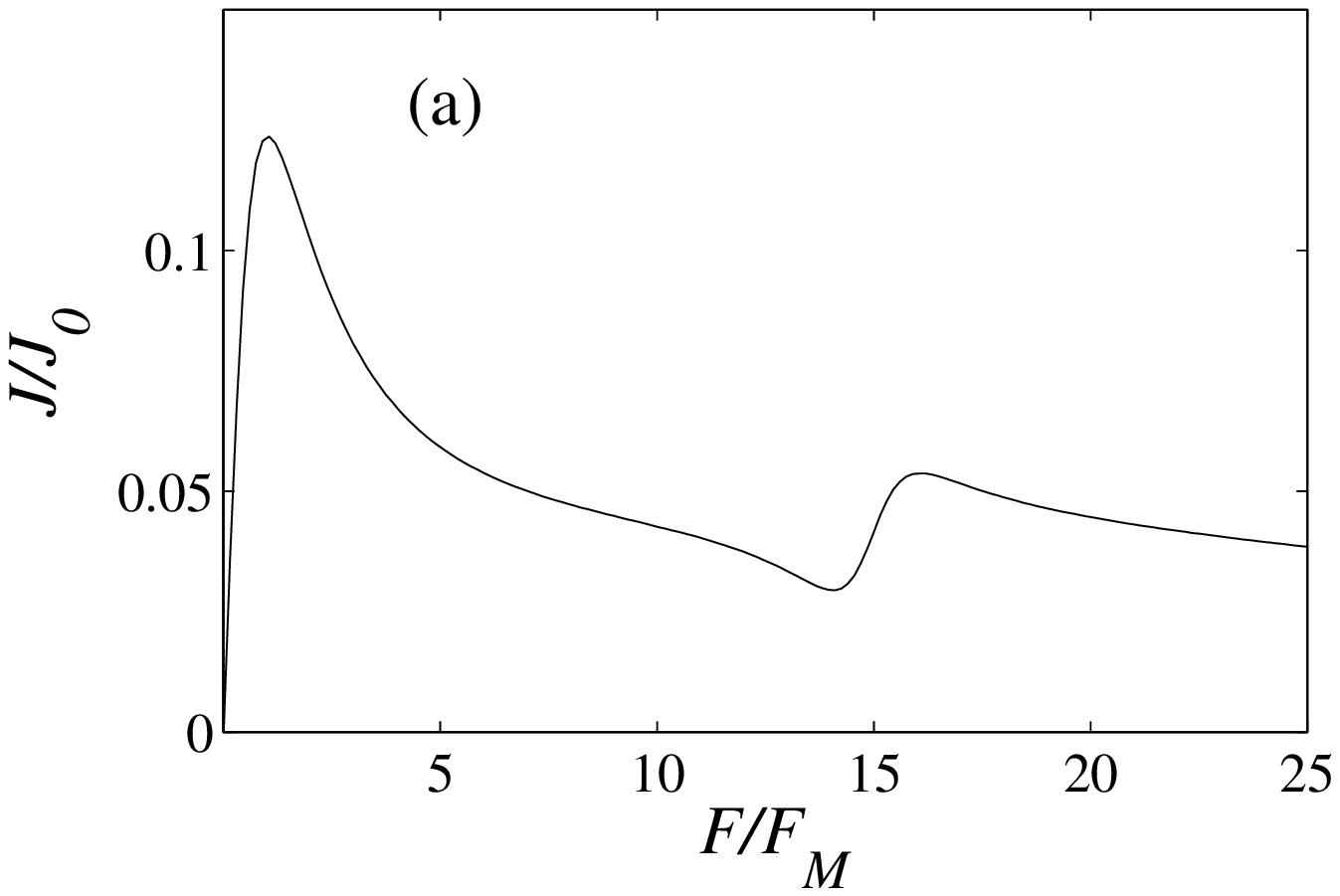}%%
\includegraphics[width=5.5cm,angle=0]{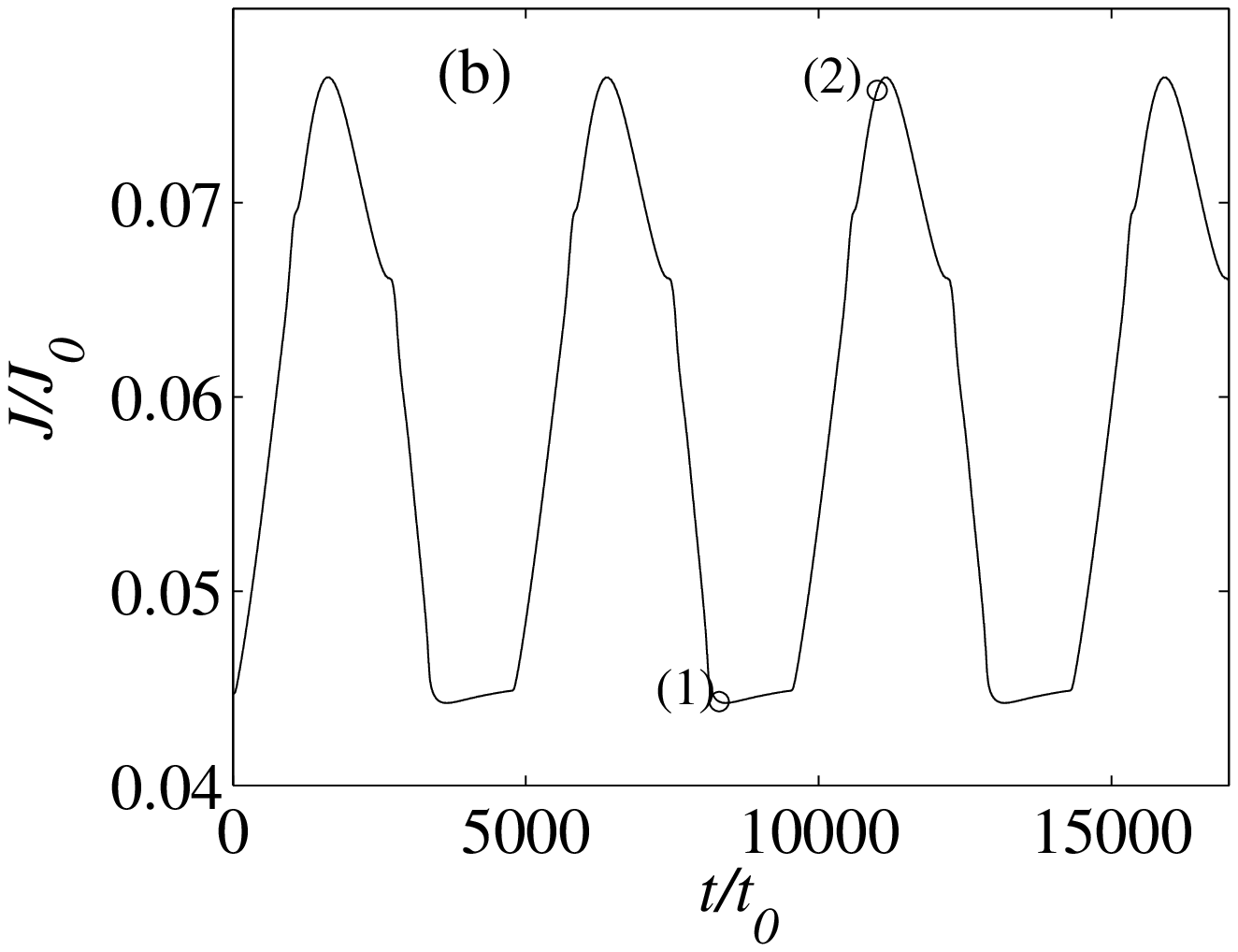}%%
\end{center}
\begin{center}
\includegraphics[width=5.5cm,angle=0]{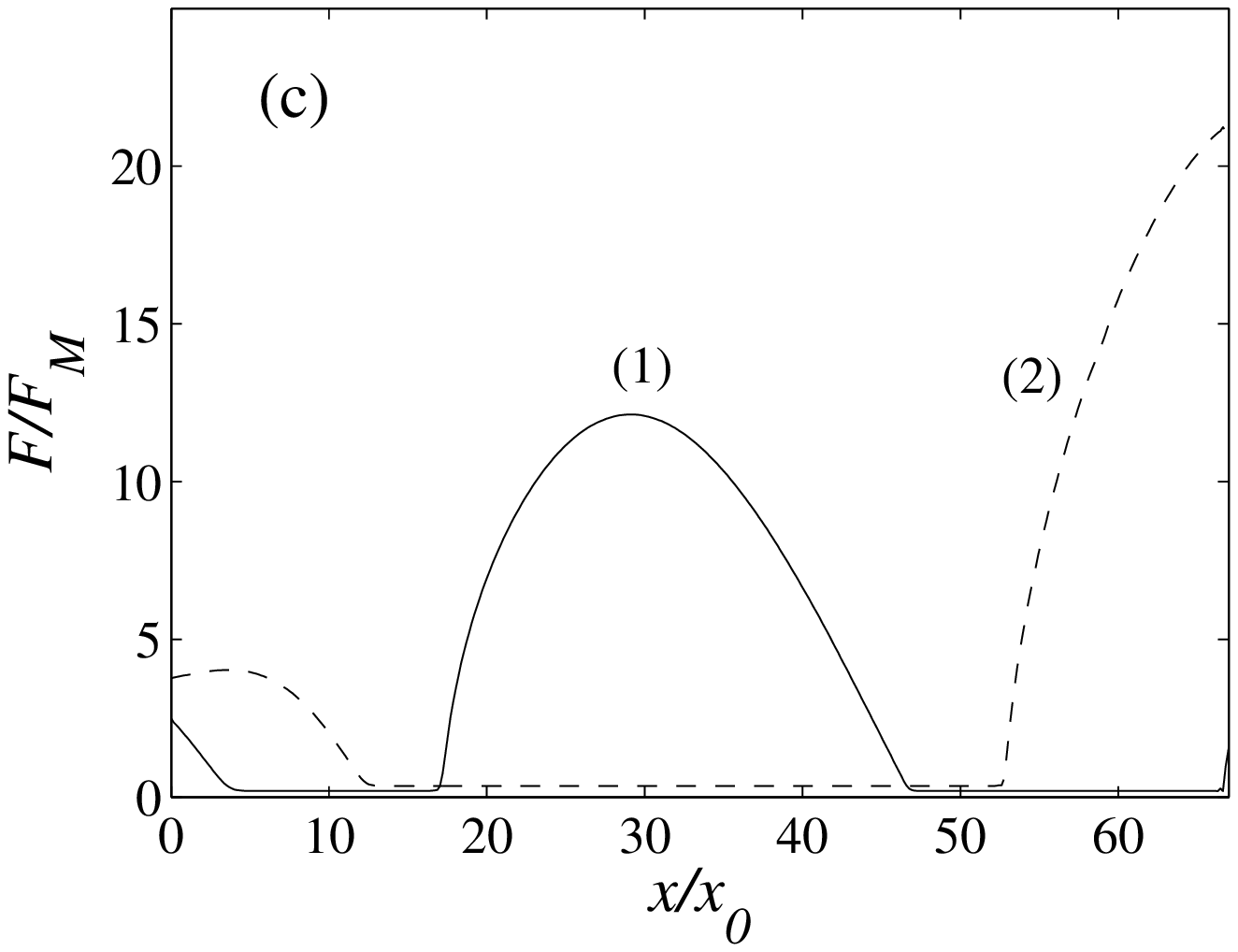}%%
\includegraphics[width=5.5cm,angle=0]{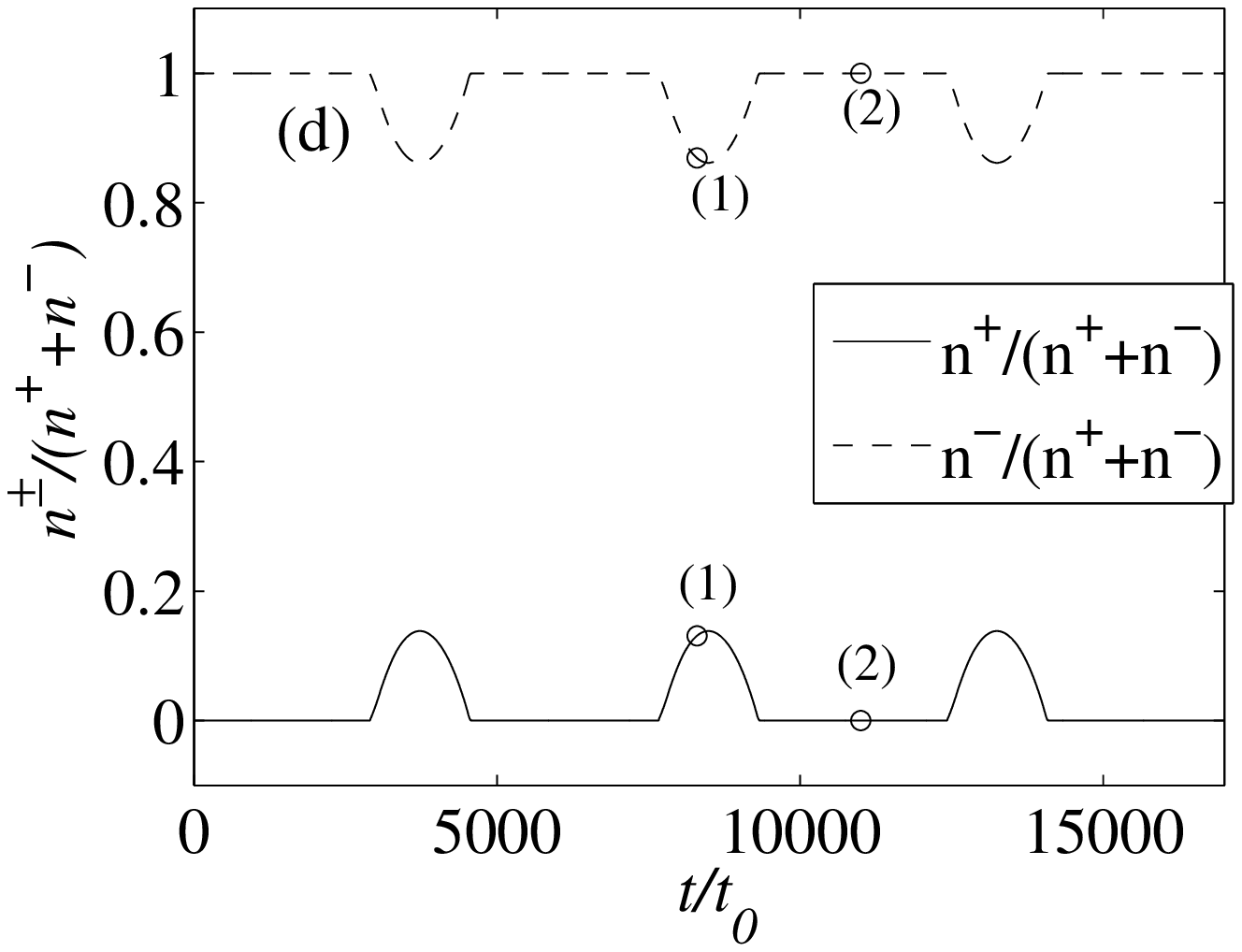}%%
\vspace{0.2cm} \caption{(a) Electron current vs field in a spatially uniform stationary state.  (b) Total current density vs time.  (c) Electric
field profile at different times of one current self-oscillations cycle. At time $(1)$ the field is above the resonant value for the middle SL point $x=Nl/2$. (d) Electron densities  $n^\pm/(n^++n^-)$ vs time for point $x=Nl/2$. When the electric field is above the resonant value (time $(1)$), the electron transport between minibands occurs. } \label{fig1}
\end{center}
\end{figure}
 Figure \ref{fig1} (b) illustrates the resulting stable self-sustained current oscillations. They are due to the
 periodic formation of a pulse of the electric field at the cathode $x = 0$ and its motion through the SL.
 Figure \ref{fig1} (a) depicts the electron current vs field in a spatially uniform stationary state, with a local maximum at
 the field resonant value $2g/(el)$. Figure \ref{fig1} (c) depicts the electric field profile at different times during one self-sustained current
 cycle.  Figure \ref{fig1} (d) shows the tunneling transport between minibands when the
 electric field is above the resonant value (time $(1)$) calculated at the middle point of the SL ($x=Nl/2$).
\begin{figure}
\begin{center}
\includegraphics[width=5.5cm,angle=0]{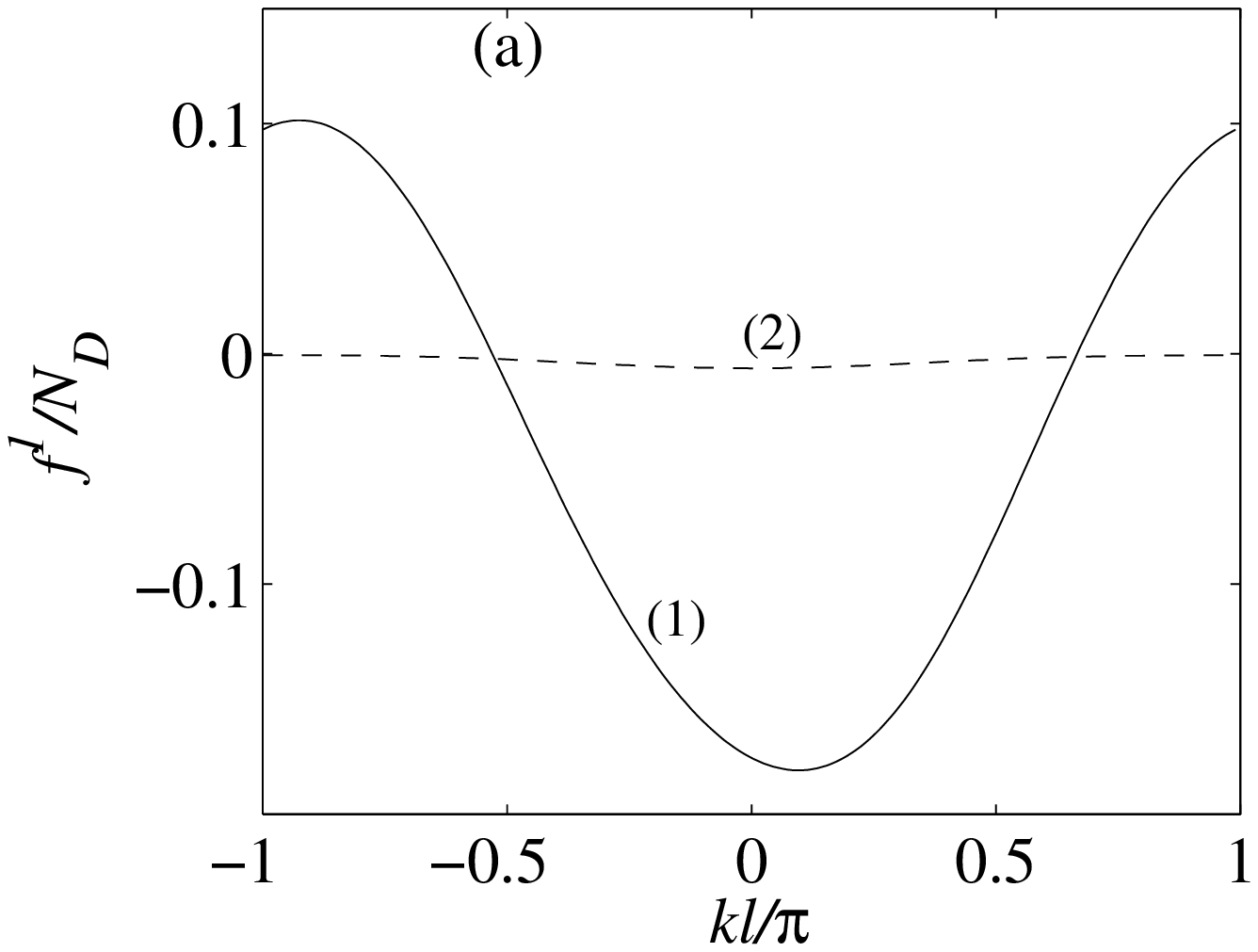}%%
\includegraphics[width=5.5cm,angle=0]{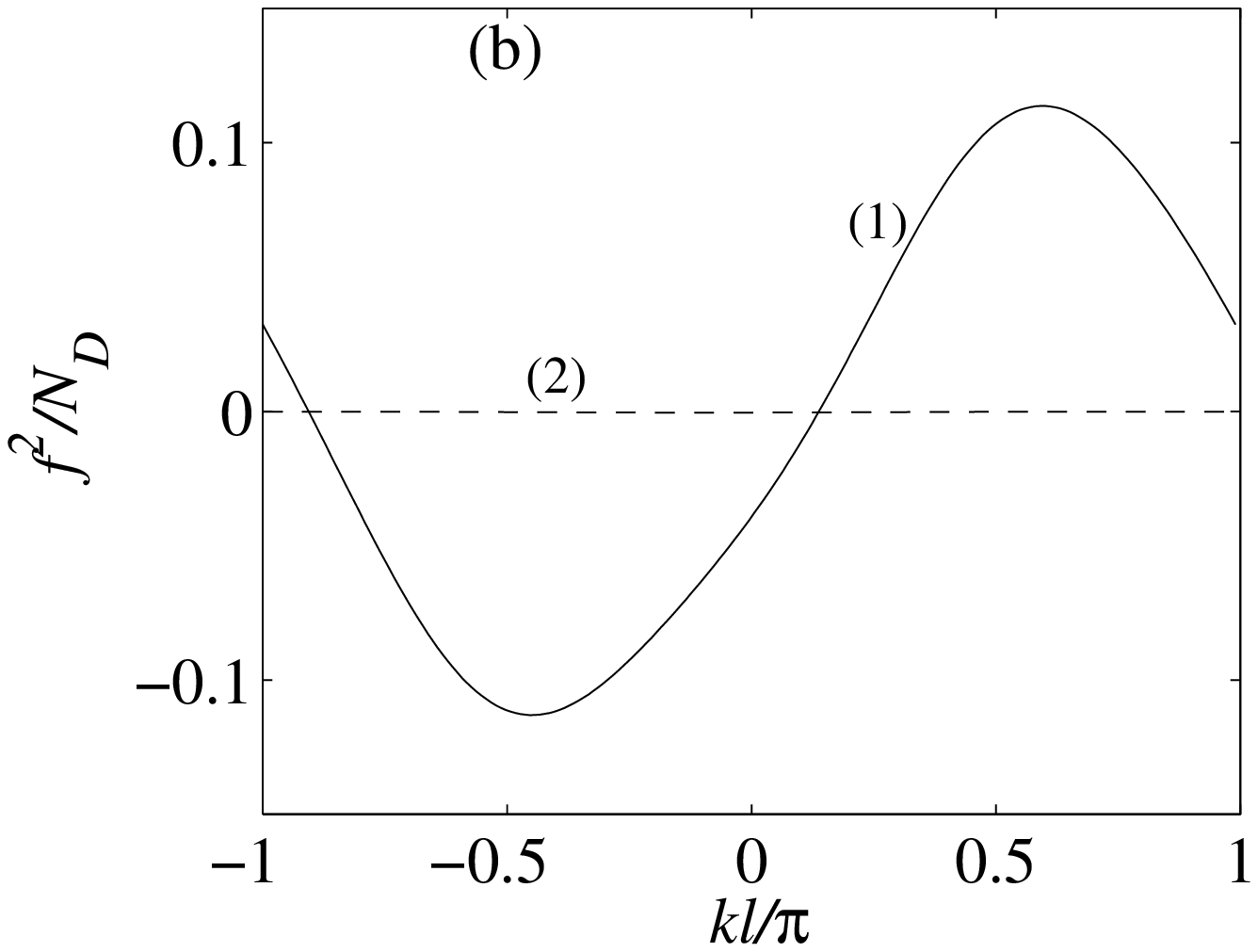}%%
\end{center}
\begin{center}
\includegraphics[width=5.5cm,angle=0]{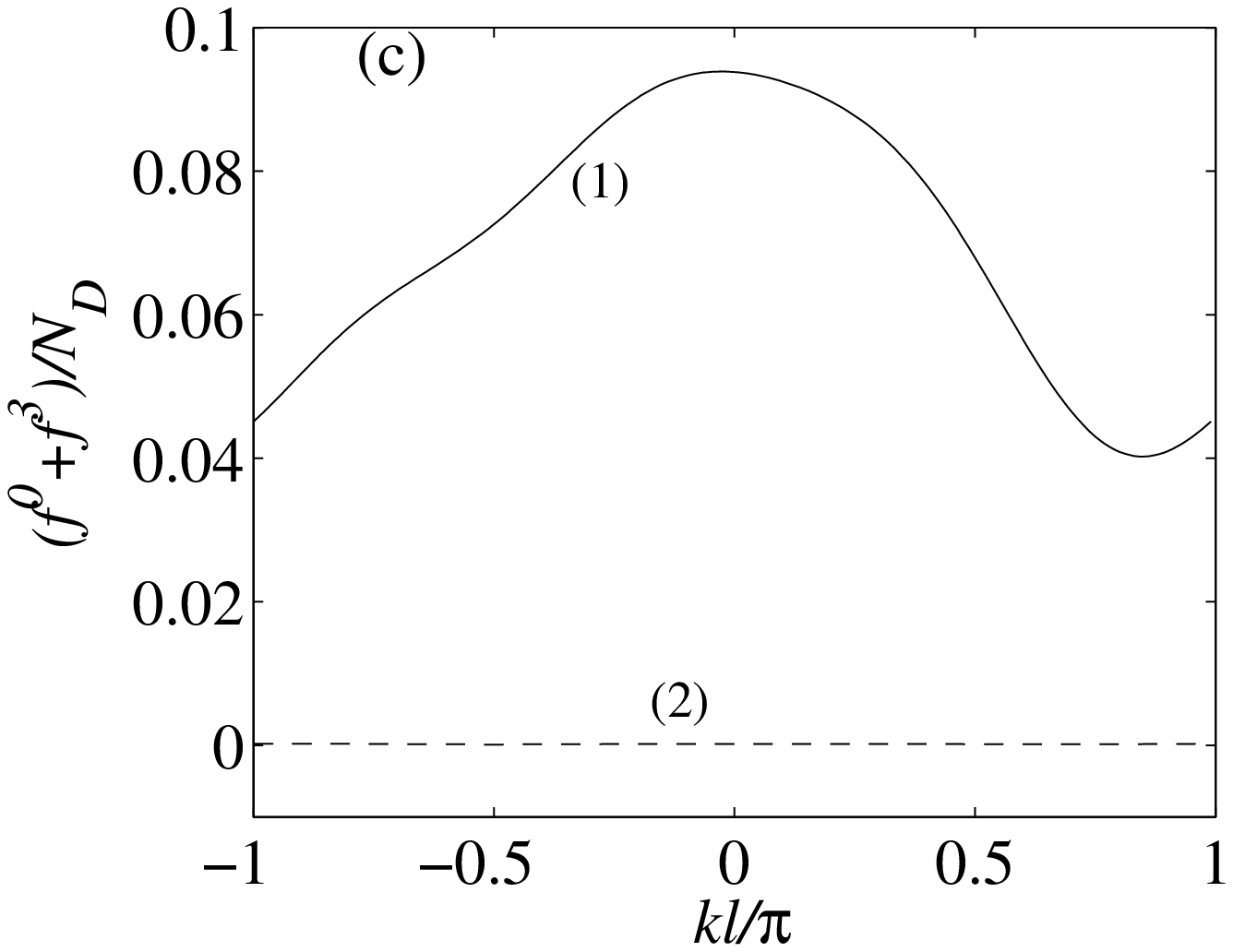}%%
\includegraphics[width=5.5cm,angle=0]{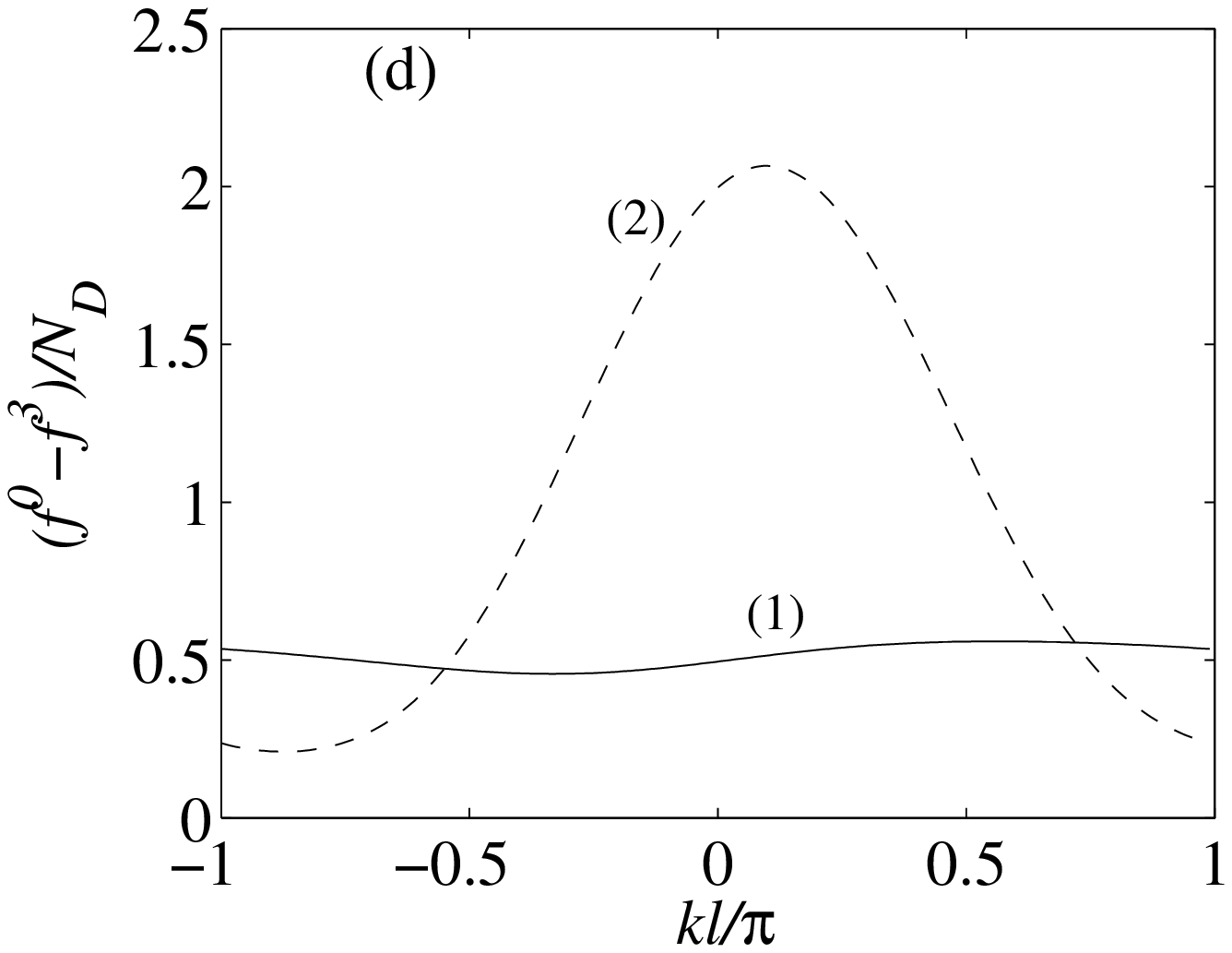}%%
\vspace{0.2cm} \caption{(a)-(b) Wigner matrix off-diagonal terms $f^1$ and $f^2$ vs $k$, at time $(1)$ (tunneling transport), and time $(2)$ (no tunneling).  (c)-(d) Wigner matrix diagonal terms $f^0 \pm f^3$ vs $k$ } \label{fig2}
\end{center}
\end{figure}

Figure \ref{fig2} shows the Wigner matrix elements $f^i$, from equations  (\ref{47})-(\ref{50}), (\ref{54})-(\ref{57}) and (\ref{fourier}), for the middle SL point ($x=Nl/2$) vs $k$ at times $(1)$ (with tunneling transport between minibands) and $(2)$ (with no  tunneling). Figure \ref{fig2}(a)-(b) illustrates the Wigner matrix off-diagonal terms $f^1$ and $f^2$, which are responsible for the tunneling transport between minibands. Figure \ref{fig2}(c)-(d) shows $f^0 \pm f^3$, which are related with the electron densities $n^{\pm}$.
\begin{figure}
\begin{center}
\includegraphics[width=5.5cm,angle=0]{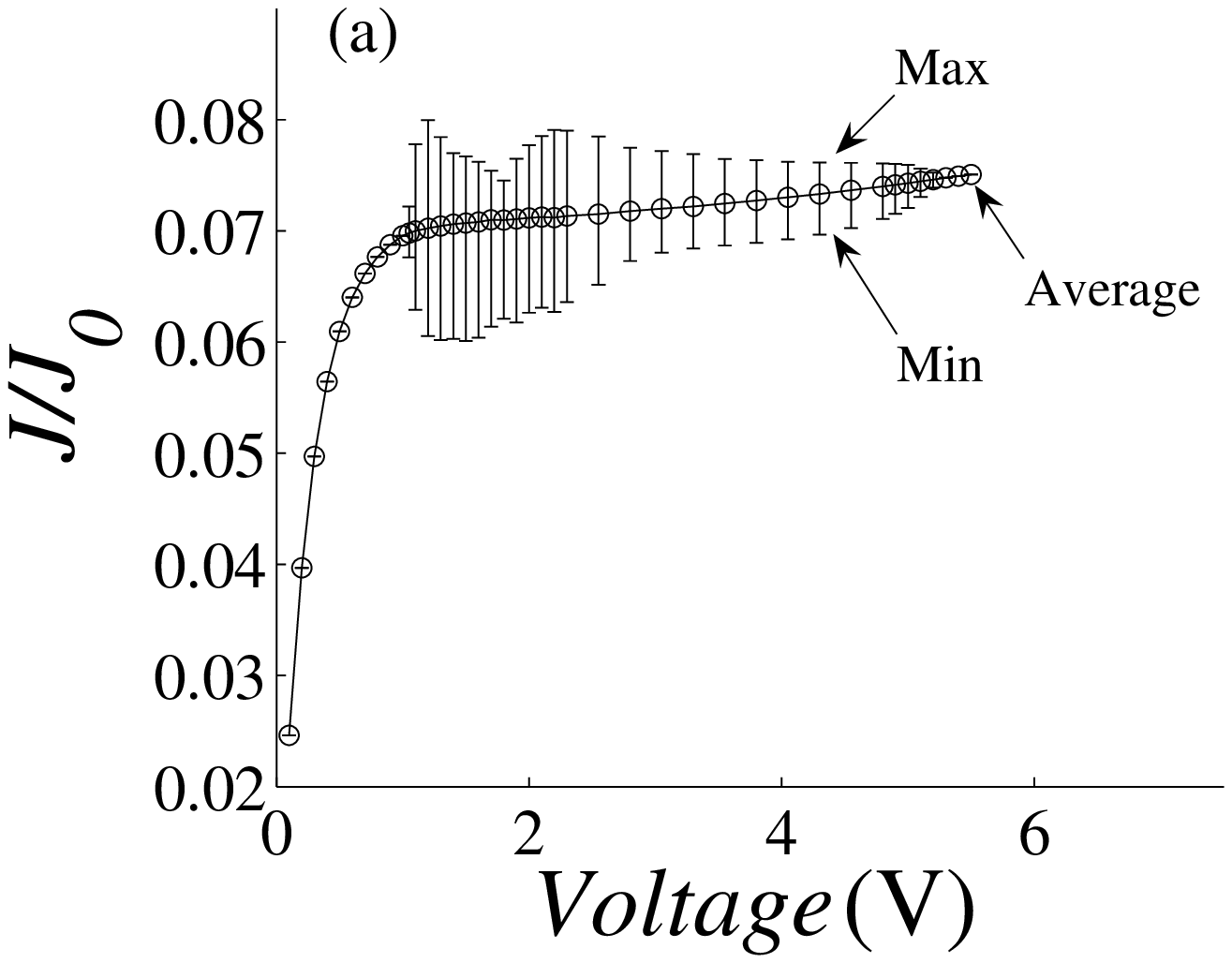}%%
\includegraphics[width=5.5cm,angle=0]{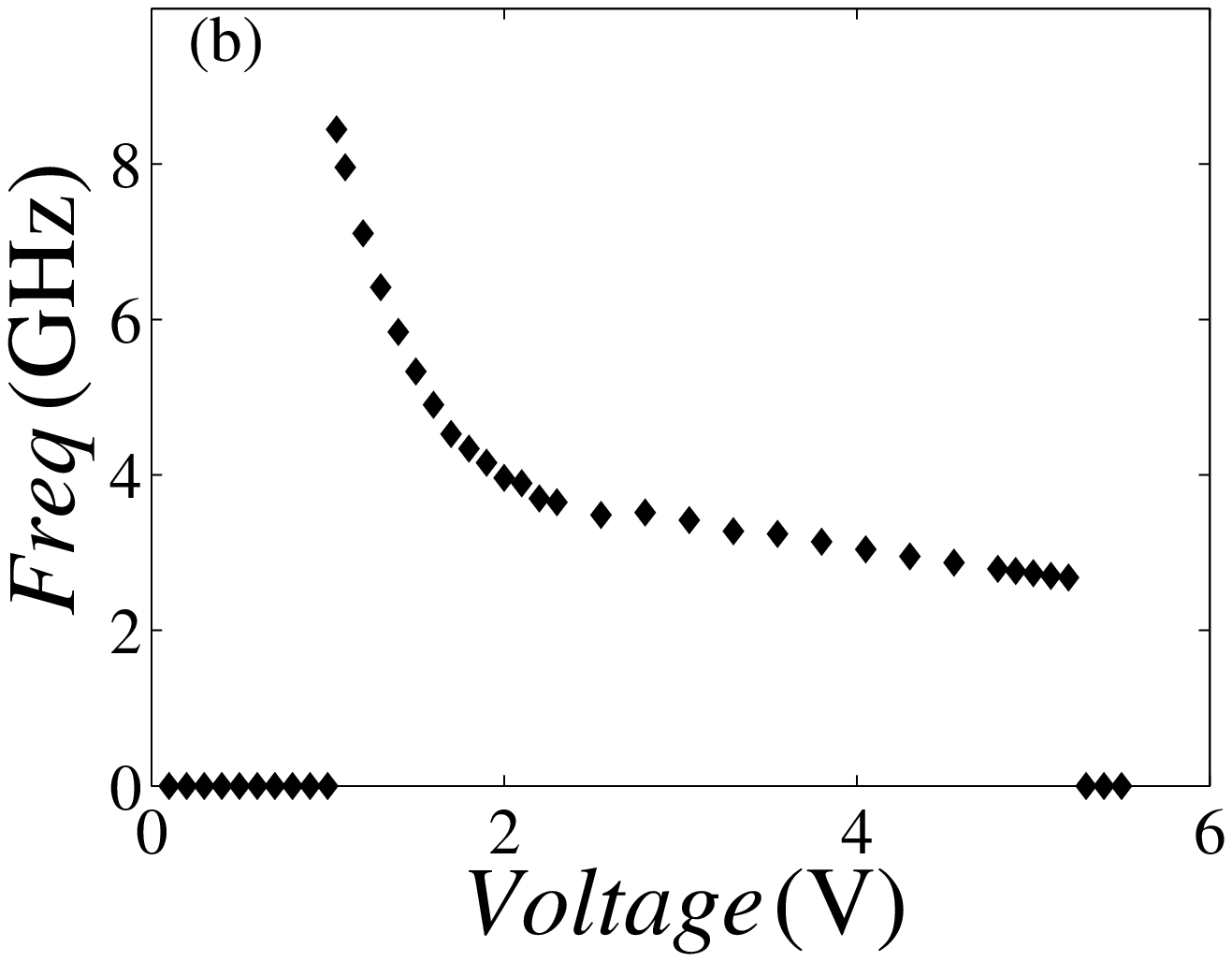}%%
\vspace{0.2cm} \caption{(a) Total current density (average, maximum and minimum values) vs voltage bias.  (b) Current oscillation frequencies vs voltage bias. } \label{fig3}
\end{center}
\end{figure}

Figure \ref{fig3} illustrates the effect of varying the voltage bias on the  total current for a $N=60$ period SL. Figure \ref{fig3} (a) depicts the total current density average, maximum and minimum values for different voltages. It can be seen that when the bias is above a critical voltage, the current self sustained oscillations appear and their amplitude increases from zero at the bifurcation point. This circumstance does not depend on whether the voltage is increasing or decreasing, therefore the critical voltage corresponds to a supercritical Hopf bifurcation. Figure \ref{fig3} (b) shows that the oscillation frequencies decrease as the voltage increases above its critical value. Immediately above the critical voltage, self-oscillations are due to repeated triggering of small pulses of the electric field that die near the cathode and before they can reach the end of the SL. As the voltage increases, the pulses are able to grow and reach the anode region. Since their average velocity does not vary that much, the oscillation frequency is correspondingly smaller. In a transition region between 1.5 and 3V, the current oscillation is somewhat irregular. The region of self-oscillations ends at a larger voltage of about 5.3V. Similar phenomena are observed in models of the Gunn effect in bulk GaAs. See Chapter 6 in Ref. \onlinecite{BT10}.

If we use parameters corresponding to a weakly coupled SL with miniband widths below 1 meV (that come from using wider quantum barriers), we run into problems of numerical convergence and, possibly, breakdown of the Chapman-Enskog perturbation scheme. To explore the limit of weakly coupled SL, a different perturbation scheme based on miniband smallness seems necessary. This is outside the scope of the present paper.

\section{Conclusions}
\label{sec:5}
For strongly coupled SLs having two populated minibands, we have introduced a kp Hamiltonian that contains a field-dependent tunneling term and derived the
corresponding Wigner-Poisson-BGK system of equations. The collision model comprises two terms, a BGK term trying to bring the Wigner matrix closer to
a broadened Fermi-Dirac local equilibrium at each miniband, and a scattering term that brings down electrons from the upper to the lower miniband. By
using the Chapman-Enskog method, we have derived quantum drift-diffusion equations for the miniband populations which contain
generation-recombination terms. As it should be, the recombination terms vanish if there is no inter-miniband scattering and the off-diagonal terms
in the Hamiltonian are zero. These terms represent miniband coupling due to the electric field and originate the resonant tunneling transport.
For a superlattice under dc voltage bias in the growth direction, numerical solutions of the corresponding quantum drift-diffusion equations show
self-sustained current oscillations due to periodic recycling and motion of electric field pulses, and resonant tunneling between minibands when the electric
field is above the resonant value. Numerical reconstruction of the Wigner functions during self-oscillations confirms this picture.

\acknowledgments
This research has been supported by the Spanish Ministerio de Ciencia e Innovaci\'on (MICINN) through Grant No. FIS2008-04921-C02-01.

\appendix
\renewcommand{\theequation}{A.\arabic{equation}}
\section{Detailed expressions for $D_{\pm}$ and $R$}
\label{appA} The recombination term $R(n^+,n^-,F)$ (\ref{61}) depends on $\varphi_0^2$ and $\psi_0^2$ which can be obtained from  (\ref{49}) and
(\ref{56}) for $j=0$, taking into account that $\Delta_0^-\mathcal F =0$, $\Delta_0^+\mathcal F = 2 \mathcal F $ and $\phi_0^\pm = n^\pm$:  \\
\begin{eqnarray}
 \varphi_0^2 &=& {\delta \mathcal F\, (n^+-n^-) \over 1+\eta_1^2 + 4\delta^2  \mathcal F^2} \nonumber \\
 \psi_0^2  &=& {1 \over 1+\eta_1^2 + 4\delta^2  \mathcal F^2} \left[ {\alpha\tau \over \hbar}\left[ {\delta(1+\eta_1^2- 4\delta^2  \mathcal F^2)
 (1-\eta_1^2- 4\delta^2  \mathcal F^2(n^+-n^-)) \over  1+\eta_1^2 + 4\delta^2  \mathcal F^2}\left. {\partial \mathcal F \over \partial t}\right|_0 -
 \right.\right.      \nonumber \\
 && {4\delta^3\mathcal F^2 \over 1+\eta_1^2 + 4\delta^2  \mathcal F^2}\left( 2(n^+-n^-)\left. {\partial \mathcal F \over \partial t}\right|_0 +
\mathcal F\, \left(\left. {\partial n^+ \over \partial t}\right|_0 -  \left. {\partial n^- \over \partial t}\right|_0 \right)\right) + \nonumber \\
&& \left. \delta\,\left((n^+-n^-)\left. {\partial \mathcal F \over \partial t}\right|_0 + \mathcal F\left( \left. {\partial n^+ \over \partial
t}\right|_0 - \left. {\partial n^- \over \partial t}\right|_0 \right)\right) \right] -  \nonumber \\
&& \eta_2\delta \mathcal F \left( {1-\eta_1^2-4\delta^2 \mathcal F^2 (n^+-n^-) \over 1+\eta_1^2 + 4\delta^2  \mathcal F^2} + 2n^+ \right)
- \nonumber \\
&& {\alpha\tau\over\hbar}\Delta^-\left[\mbox{Im} \varphi_1^2 - \eta_1 \mbox{Im} \varphi_1^1 + 2\delta \mathcal F \mbox{Im} \varphi_1^3\right] -
{\gamma\tau\over\hbar}\left(-{4\eta_1\delta\mathcal F (n^+-n^-)\over 1+\eta_1^2 + 4\delta^2  \mathcal F^2} + \right.      \nonumber \\
&& \left. 2\delta\mathcal F \Delta^- \mbox{Im} \varphi_1^0 + \Delta^+(\mbox{Re} \varphi_1^1 + \eta_1 \mbox{Re} \varphi_1^2) \right) \nonumber
\end{eqnarray}

The time derivatives $\displaystyle\left.{\partial n^\pm \over\partial t} \right|_0$ and $\displaystyle\left. {\partial F \over \partial
t}\right|_0$, are obtained from the first two terms of the Chapman-Enskog expansion of (\ref{58}) and (\ref{59}) respectively:
\begin{eqnarray}
\left.{\partial n^\pm \over \partial t}\right|_0 &=& \mp Q^{(0)} - \Delta^-D_\pm^{(0)} = \mp{n^+
\over \tau_{sc}} \mp {2\delta^2 \mathcal F^2 (n^+-n^-) \over \tau (1+\eta_1^2+4\delta^2\mathcal F^2)} +  \nonumber \\
&& {\alpha \pm \gamma \over \hbar}\Delta^-\left[\phi_1^{\pm}\left({\vartheta_1 \over 1 + \vartheta_1^2} +
\eta_1\delta^2 \left(\mp \Delta^-\mathcal F\Delta^+\mathcal F \mbox{Re} Z_1 + \right. \right. \right. \nonumber \\
&&  \mbox{Re} Z_1((\Delta^-\mathcal F)^2 \mbox{Im} M_1^+ + (\Delta^+\mathcal F)^2 \mbox{Im} M_1^-) + \nonumber \\
&& \left. \left. \left. \mbox{Im} Z_1((\Delta^-\mathcal F)^2 \mbox{Re} M_1^+ + (\Delta^+\mathcal F)^2 \mbox{Re} M_1^-)\right) \right) + \right. \nonumber \\
&& \phi_1^{\mp}\eta_1 \delta^2[\mbox{Re} Z_1((\Delta^-\mathcal F)^2 \mbox{Im}
M_1^+ - (\Delta^+\mathcal F)^2 \mbox{Im} M_1^-) + \nonumber \\
&& \left. \mbox{Im} Z_1((\Delta^-\mathcal F)^2 \mbox{Re} M_1^+ - (\Delta^+\mathcal F)^2 \mbox{Re} M_1^-)] \right] \nonumber \\
\left. \varepsilon{\partial F \over \partial t}\right|_0 &=& J -  e\left<D_+^{(0)}+D_-^{(0)}\right> = J -
{e\alpha \over \hbar}\left[(\phi_1^++\phi_1^-)\left({-\vartheta_1 \over 1 + \vartheta_1^2} \right.\right. - \nonumber \\
&&  \eta_1 \delta^2 (\Delta^- \mathcal F)^2 (\mbox{Re} M_1^+ \mbox{Im} Z_1 + \mbox{Im} M_1^+ \mbox{Re} Z_1) +  \nonumber \\
&& \left. {\gamma\over\alpha}\eta_1 \delta^2 \Delta^+ \mathcal F \Delta^- \mathcal F  \mbox{Re} Z_1\right) +
(\phi_1^+-\phi_1^-)\left(\eta_1\delta^2\Delta^+\mathcal F \Delta^-\mathcal F \mbox{Re} Z_1 -    \right.    \nonumber \\
&& \left.\left. {\gamma\over\alpha}\left({\vartheta_1 \over 1+ \vartheta_1^2} + \eta_1\delta^2(\Delta^+ \mathcal F)^2 (\mbox{Re} M_1^- \mbox{Im} Z_1 + \mbox{Im} M_1^-
\mbox{Re} Z_1)\right)\right)\right] \nonumber
\end{eqnarray}

The expression of $D_{\pm}(n^+,n^-,F)$ is based on the first two terms of the Chapman-Enskog expansion $D_{\pm}^{(0)}$ and $D_{\pm}^{(1)}$:

\begin{eqnarray}
 D_{\pm}(n^+,n^-,F) &=& D_{\pm}^{(0)}(n^+,n^-,F) + D_{\pm}^{(1)}(n^+,n^-,F)  \nonumber
\end{eqnarray}
Where $D_{\pm}^{(0)}$ and $D_{\pm}^{(1)}$ are as follows:
\begin{eqnarray}
 D_{\pm}^{(0)}(n^+,n^-,F) &=& {\alpha \pm \gamma \over \hbar} \mbox{Im}(\varphi_1^0 \pm \varphi_1^3) = \nonumber \\
 && -{\alpha \pm \gamma \over \hbar}\left[ \phi_1^{\pm}\left({\vartheta_1 \over 1+\vartheta_1^2} \mp \eta_1
\delta^2 (\Delta^- \mathcal F)(\Delta^+ \mathcal F)\mbox{Re} Z_1  \right. \right. \nonumber \\
&& \left. \left. + {\eta_1 \delta^2 \over 2}((\mbox{Im} M_1^+(\Delta^- \mathcal F)^2 \nonumber + \mbox{Im} M_1^-(\Delta^+ \mathcal
F)^2)\mbox{Re} Z_1 \right. \right. \nonumber \\
&& \left.\left. + (\mbox{Re} M_1^+(\Delta^- \mathcal F)^2 + \mbox{Re} M_1^-(\Delta^+ \mathcal F)^2)\mbox{Im} Z_1) \right) \right.\nonumber\\
&& \left. + \,\phi_1^{\mp}\,{\eta_1 \delta^2 \over 2}( \mbox{Im} M_1^+(\Delta^- \mathcal F)^2 - \mbox{Im} M_1^-(\Delta^+ \mathcal F)^2 )\mbox{Re} Z_1\right. \nonumber \\
&& \left. + \,(\mbox{Re} M_1^+(\Delta^- \mathcal F)^2 - \mbox{Re} M_1^-(\Delta^+ \mathcal F)^2)\mbox{Im} Z_1) \right] \nonumber
\end{eqnarray}
\begin{eqnarray}
 D_{\pm}^{(1)}(n^+,n^-,F) &=& {\alpha \pm \gamma \over \hbar} \mbox{Im}(\psi_1^0 \pm \psi_1^3) = \nonumber\\
&& {\alpha \pm \gamma \over \hbar} \left[\mbox{Re} S_1^0 \, \mbox{Im} A_1^{\pm} + \mbox{Im} S_1^0\, \mbox{Re} A_1^{\pm} \pm \mbox{Re} S_1^3\, \mbox{Im}
C_1^{\pm} \pm \mbox{Im} S_1^3\, \mbox{Re} C_1^{\pm} \right. \nonumber \\
&&\left. + \eta_1 \delta (\mbox{Re} Z_1\, \mbox{Im} B_1^{\pm} + \mbox{Re} Z_1\, \vartheta_1\, \mbox{Re} B_1^{\pm} + \mbox{Im} Z_1\, \mbox{Re} B_1^{\pm} - \mbox{Im} Z_1\, \vartheta_1\, \mbox{Im}
B_1^{\pm}) \right] \nonumber
\end{eqnarray}
The functionals $S_1(n^+,n^-,F)$, $A_1^\pm(F)$, $B_1^\pm(F)$ and $C_1^\pm(F)$ are as follows:
\begin{eqnarray}
\mbox{Re} A_1^{\pm} &=& {1\over 1+\vartheta_1^2} - \eta_1 \delta^2 \Delta^- \mathcal F\, (\mbox{Re} Z_1\, \mbox{Re} M_1^+ \Delta^-
\mathcal F - \mbox{Im} Z_1 ( \Delta^- \mathcal F\, \mbox{Im} M_1^+ \mp \Delta^+ \mathcal F)), \nonumber \\
\mbox{Im} A_1^{\pm} &=& {-\vartheta_1 \over 1+\vartheta_1^2} - \eta_1 \delta^2 \Delta^- \mathcal F (\mbox{Re} Z_1 (\Delta^-
\mathcal F \, \mbox{Im} M_1^+ \mp \Delta^+ \mathcal F)+ \mbox{Im} Z_1 \, \Delta^- \mathcal F \, \mbox{Re} M_1^+), \nonumber \\
\mbox{Re} B_1^{\pm} &=& \mbox{Re} S_1^1 (-\Delta^- \mathcal F \mbox{Im} M_1^+ \pm \Delta^+ \mathcal F)- \mbox{Im} S_1^1 \, \Delta^- \mathcal F \,\mbox{Re}
M_1^+ \mp \mbox{Re} S_1^2 \, \Delta^+ \mathcal F \, \mbox{Re} M_1^- \nonumber \\
&& - \mbox{Im} S_1^2(\Delta^- \mathcal F \mp \Delta^+ \mathcal F \, \mbox{Im} M_1^-), \nonumber \\
\mbox{Im} B_1^{\pm} &=& \mbox{Re} S_1^1 \, \Delta^- \mathcal F \, \mbox{Re} M_1^+ + \mbox{Im} S_1^1 (\pm\Delta^+\mathcal F - \Delta^- \mathcal F \mbox{Im}
M_1^+), \nonumber \\
&& + \mbox{Re} S_1^2(\Delta^- \mathcal F \mp \Delta^+ \mathcal F \mbox{Im} M_1^-) \mp \mbox{Im} S_1^2\, \Delta^+ \mathcal F\, \mbox{Re} M_1^-
\nonumber \\
\mbox{Re} C_1^{\pm} &=& {1 \over 1+ \vartheta_1^2} - \eta_1 \delta^2 \Delta^+ \mathcal F (\mbox{Re} Z_1 \,\Delta^+ \mathcal F \, \mbox{Re}
M_1^- - \mbox{Im} Z_1 ( \Delta^+ \mathcal F\, \mbox{Im} M_1^- \mp \Delta^- \mathcal F)), \nonumber \\
\mbox{Im} C_1^{\pm} &=& {-\vartheta_1 \over 1+\vartheta_1^2} - \eta_1\delta^2 \Delta^+ \mathcal F (\mbox{Re} Z_1 (
\Delta^+ \mathcal F \mbox{Im} M_1^- \mp \Delta^- \mathcal F) + \mbox{Im} Z_1 \, \Delta^+ \mathcal F\, \mbox{Re} M_1^-) \nonumber \\
\mbox{Re} S_1^0 &=& -\tau\, \left. {\partial \mbox{Re} \varphi_1^0 \over \partial t}\right|_0 - {\alpha \tau \over 2
\hbar}\Delta^-\mbox{Im} \varphi_2^0 - {\gamma \tau \over 2 \hbar}\Delta^-\mbox{Im} \varphi_2^3 \nonumber \\
\mbox{Im} S_1^0 &=& -\tau\, \left. {\partial \mbox{Im} \varphi_1^0 \over
\partial t}\right|_0 + {\alpha \tau \over 2 \hbar}\Delta^-(\mbox{Re} \varphi_2^0 - \varphi_0^0) + {\gamma \tau \over 2 \hbar}\Delta^-(\mbox{Re} \varphi_2^3 - \varphi_0^3) \nonumber \\
\mbox{Re} S_1^1 &=& -\tau\, \left. {\partial \mbox{Re} \varphi_1^1 \over \partial t}\right|_0 - \eta_2 \mbox{Re}\varphi_1^1 - {\alpha \tau \over 2\hbar}\Delta^-\mbox{Im}
\varphi_2^1 - {2\gamma\tau\over\hbar}\mbox{Re} \varphi_1^2 + {\gamma \tau \over 2 \hbar}\Delta^+(\mbox{Re} \varphi_2^2+\varphi_0^2) \nonumber \\
\mbox{Im} S_1^1 &=& -\tau\, \left. {\partial \mbox{Re} \varphi_1^1 \over \partial t}\right|_0 - \eta_2 \mbox{Im}\varphi_1^1 +{\alpha \tau \over 2\hbar}\Delta^-(\mbox{Re}
\varphi_2^1-\varphi_0^1) - {2\gamma\tau\over\hbar}\mbox{Im} \varphi_1^2 + {\gamma \tau \over 2 \hbar}\Delta^+\mbox{Im} \varphi_2^2 \nonumber \\
\mbox{Re} S_1^2 &=& -\tau\, \left. {\partial \mbox{Re} \varphi_1^2 \over \partial t}\right|_0 - \eta_2 \mbox{Re}\varphi_1^2 - {\alpha \tau \over 2\hbar}\Delta^-\mbox{Im}
\varphi_2^2 + {2\gamma\tau\over\hbar}\mbox{Re} \varphi_1^1 - {\gamma \tau \over 2 \hbar}\Delta^+(\mbox{Re} \varphi_2^1+\varphi_0^1) \nonumber \\
\mbox{Im} S_1^2 &=& -\tau\, \left. {\partial \mbox{Re} \varphi_1^2 \over \partial t}\right|_0 - \eta_2 \mbox{Im}\varphi_1^2 +{\alpha \tau \over 2\hbar}\Delta^-(\mbox{Re}
\varphi_2^2-\varphi_0^2) + {2\gamma\tau\over\hbar}\mbox{Im} \varphi_1^1 - {\gamma \tau \over 2 \hbar}\Delta^+\mbox{Im} \varphi_2^1 \nonumber \\
\mbox{Re} S_1^3 &=& -\tau\, \left. {\partial \mbox{Re} \varphi_1^3 \over \partial t}\right|_0 - \eta_2( \mbox{Re}\varphi_1^3 + \mbox{Re} \varphi_1^0) - {\alpha \tau \over
2\hbar}\Delta^-\mbox{Im} \varphi_2^3 - {\gamma \tau \over 2 \hbar}\Delta^-\mbox{Im} \varphi_2^0 \nonumber \\
\mbox{Im} S_1^3 &=& -\tau\, \left. {\partial \mbox{Im} \varphi_1^3 \over \partial t}\right|_0 - \eta_2( \mbox{Im}\varphi_1^3 + \mbox{Im} \varphi_1^0) + {\alpha \tau \over
2\hbar}\Delta^-(\mbox{Re} \varphi_2^3-\varphi_0^3) + {\gamma \tau \over 2 \hbar}\Delta^-(\mbox{Re} \varphi_2^0 - \varphi_0^0) \nonumber
\end{eqnarray}
Where the real and imaginary parts of $\varphi_j$ are:
\begin{eqnarray}
\mbox{Re} \varphi_j^0 &=& {\phi_j^++\phi_j^- \over 2}\left[{1\over 1+
\vartheta_j^2}-\eta_1\delta^2(\Delta_j^-\mathcal F)^2(\mbox{Re} M_j^+ \mbox{Re} Z_j - \mbox{Im} M_j^+ \mbox{Im} Z_j)\right] - \nonumber \\
&& {\eta_1\delta^2 \over 2}(\phi_j^+-\phi_j^-)\Delta_j^+\mathcal F \Delta_j^-\mathcal F \mbox{Im} Z_j \nonumber \\
\mbox{Im} \varphi_j^0 &=& {\phi_j^++\phi_j^- \over 2}\left[{-\vartheta_j\over 1+ \vartheta_j^2}-
\eta_1\delta^2(\Delta_j^-\mathcal F)^2(\mbox{Re} M_j^+ \mbox{Im} Z_j + \mbox{Im} M_j^+ \mbox{Re} Z_j)\right] + \nonumber \\
&& {\eta_1\delta^2 \over 2}(\phi_j^+-\phi_j^-)\Delta_j^+\mathcal F \Delta_j^-\mathcal F \mbox{Re} Z_j \nonumber \\
\mbox{Re} \varphi_j^1 &=& {\eta_1\delta \over 2}\left[-(\phi_j^++\phi_j^-)\Delta_j^-\mathcal F(\mbox{Im} M_j^+ (\mbox{Re}
Z_j-\vartheta_j \mbox{Im} Z_j)\right. + \nonumber \\
&& \mbox{Re} M_j^+ (\mbox{Im} Z_j+\vartheta_j \mbox{Re} Z_j)) + \left.
(\phi_j^+-\phi_j^-)\Delta_j^+\mathcal F(\mbox{Re} Z_j - \vartheta_j \mbox{Im} Z_j)\right] \nonumber \\
\mbox{Im} \varphi_j^1 &=& {\eta_1\delta \over 2}\left[(\phi_j^++\phi_j^-)\Delta_j^-\mathcal F(\mbox{Re} M_j^+ (\mbox{Re}
Z_j-\vartheta_j \mbox{Im} Z_j)\right. - \nonumber \\
&& \mbox{Im} M_j^+ (\mbox{Im} Z_j+\vartheta_j \mbox{Re} Z_j)) + \left.
(\phi_j^+-\phi_j^-)\Delta_j^+\mathcal F(\mbox{Im} Z_j + \vartheta_j \mbox{Re} Z_j)\right] \nonumber \\
\mbox{Re} \varphi_j^2 &=& {\eta_1\delta \over 2}\left[(\phi_j^++\phi_j^-)\Delta_j^-\mathcal F(\mbox{Im} Z_j+\vartheta_j \mbox{Re} Z_j)\right. + \nonumber \\
&& \left. (\phi_j^+-\phi_j^-)\Delta_j^+\mathcal F(\mbox{Re} M_j^- (\mbox{Re} Z_j - \vartheta_j \mbox{Im} Z_j) -
\mbox{Im} M_j^-(\mbox{Im} Z_j + \vartheta_j \mbox{Re} Z_j))\right] \nonumber \\
\mbox{Im} \varphi_j^2 &=& {\eta_1\delta \over 2}\left[(\phi_j^++\phi_j^-)\Delta_j^-\mathcal F(-\mbox{Re} Z_j+\vartheta_j \mbox{Im} Z_j)\right. + \nonumber \\
&& \left. (\phi_j^+-\phi_j^-)\Delta_j^+\mathcal F(\mbox{Re} M_j^- (\mbox{Im} Z_j + \vartheta_j \mbox{Re} Z_j) +
\mbox{Im} M_j^-(\mbox{Re} Z_j - \vartheta_j \mbox{Im} Z_j))\right] \nonumber \\
\mbox{Re} \varphi_j^3 &=& {\phi_j^+-\phi_j^- \over 2}\left[{1\over 1+
\vartheta_j^2}-\eta_1\delta^2(\Delta_j^+\mathcal F)^2(\mbox{Re} M_j^- \mbox{Re} Z_j - \mbox{Im} M_j^- \mbox{Im} Z_j)\right] - \nonumber \\
&& {\eta_1\delta^2 \over 2}(\phi_j^++\phi_j^-)\Delta_j^+\mathcal F \Delta_j^-\mathcal F \mbox{Im} Z_j \nonumber \\
\mbox{Im} \varphi_j^3 &=& {\phi_j^+-\phi_j^- \over 2}\left[{-\vartheta_j\over 1+ \vartheta_j^2}-
\eta_1\delta^2(\Delta_j^+\mathcal F)^2(\mbox{Re} M_j^- \mbox{Im} Z_j + \mbox{Im} M_j^- \mbox{Re} Z_j)\right] + \nonumber \\
&& {\eta_1\delta^2 \over 2}(\phi_j^++\phi_j^-)\Delta_j^+\mathcal F \Delta_j^-\mathcal F \mbox{Re} Z_j \nonumber
\end{eqnarray}
Now we can obtain the expressions for $\displaystyle \left. {\partial \varphi_j \over \partial t}\right|_0$:
\begin{eqnarray}
\left. {\partial \mbox{Re} \varphi_1^0 \over \partial t}\right|_0 &=& {1\over 2}\left((\phi_1^+)' \, \left.{\partial n^+
\over \partial t}\right|_0  +  (\phi_1^-)' \, \left.{\partial n^- \over \partial t}\right|_0  \right)\times \nonumber \\
&& \left[{1 \over 1+\vartheta_1^2} - \eta_1\delta^2(\Delta^- \mathcal F)^2\,(\mbox{Re} M_1^+ \mbox{Re} Z_1 - \mbox{Im} M_1^+ \mbox{Im} Z_1)\right] +\nonumber \\
&& {1\over 2}(\phi_1^+ + \phi_1^-)\left[{-2\vartheta_1 \over (1+\vartheta_1^2)^2} \left.{\partial \vartheta_1 \over \partial t}\right|_0
\right. - \nonumber \\
&&   2\eta_1\delta^2 \Delta^- \mathcal F \Delta^- \left.{\partial \mathcal F \over \partial t}\right|_0 (\mbox{Re} M_1^+ \mbox{Re} Z_1 - \mbox{Im} M_1^+ \mbox{Im} Z_1) -
 \nonumber \\
&&  \eta_1\delta^2 (\Delta^- \mathcal F)^2\left(\left.{\partial \mbox{Re} M_1^+ \over \partial t}\right|_0 \mbox{Re} Z_1 + \mbox{Re} M_1^+ \left.{\partial \mbox{Re} Z_1 \over
\partial t}\right|_0 - \mbox{Im} Z_1 \left.{\partial \mbox{Im} M_1^+ \over \partial t}\right|_0 \right. - \nonumber \\
&& \left.\left. \mbox{Im} M_1^+ \left.{\partial \mbox{Im} Z_1 \over \partial t}\right|_0\right) \right] - \nonumber \\
&& {\eta_1\delta^2 \over 2}\left(\left((\phi_1^+)' \left.{\partial n^+ \over \partial t}\right|_0 - (\phi_1^-)'
\left.{\partial n^- \over \partial t}\right|_0 \right) \Delta^+ \mathcal F \Delta^- \mathcal F \mbox{Im} Z_1 -\right. \nonumber \\
&& (\phi_1^+-\phi_1^-)\left(\Delta^+\left.{\partial \mathcal F \over \partial t}\right|_0\Delta^- \mathcal F \mbox{Im} Z_1 +
\Delta^-\left.{\partial \mathcal F \over \partial t}\right|_0\Delta^+ \mathcal F \mbox{Im} Z_1 \right. + \nonumber \\
&& \left.\left.\left.{\partial \mbox{Im} Z_1 \over \partial t}\right|_0\Delta^+ \mathcal F \Delta^- \mathcal F \right)\right) \nonumber \\
\left. {\partial \mbox{Im} \varphi_1^0 \over \partial t}\right|_0 &=& {1\over 2}\left((\phi_1^+)' \, \left.{\partial n^+
\over \partial t}\right|_0  +  (\phi_1^-)' \, \left.{\partial n^- \over \partial t}\right|_0  \right)\times \nonumber \\
&& \left[{-\vartheta_1 \over 1+\vartheta_1^2} - \eta_1\delta^2(\Delta^- \mathcal F)^2\,(\mbox{Re} M_1^+ \mbox{Im} Z_1 + \mbox{Im} M_1^+ \mbox{Re} Z_1)\right] +\nonumber \\
&& {1\over 2}(\phi_1^+ + \phi_1^-)\left[{\vartheta_1^2-1 \over (1+\vartheta_1^2)^2} \left.{\partial \vartheta_1 \over \partial t}\right|_0
\right. - \nonumber \\
&&   2\eta_1\delta^2 \Delta^- \mathcal F \Delta^- \left.{\partial \mathcal F \over \partial t}\right|_0 (\mbox{Re} M_1^+ \mbox{Im} Z_1 + \mbox{Im} M_1^+ \mbox{Re} Z_1) -
 \nonumber \\
&&  \eta_1\delta^2 (\Delta^- \mathcal F)^2\left(\left.{\partial \mbox{Re} M_1^+ \over \partial t}\right|_0 \mbox{Im} Z_1 + \mbox{Re} M_1^+ \left.{\partial \mbox{Im} Z_1 \over
\partial t}\right|_0 + \mbox{Re} Z_1 \left.{\partial \mbox{Im} M_1^+ \over \partial t}\right|_0 \right. - \nonumber \\
&& \left.\left. \mbox{Im} M_1^+ \left.{\partial \mbox{Re} Z_1 \over \partial t}\right|_0\right) \right] + \nonumber \\
&& {\eta_1\delta^2 \over 2}\left(\left((\phi_1^+)' \left.{\partial n^+ \over \partial t}\right|_0 - (\phi_1^-)'
\left.{\partial n^- \over \partial t}\right|_0 \right) \Delta^+ \mathcal F \Delta^- \mathcal F \mbox{Re} Z_1 + \right.\nonumber \\
&& (\phi_1^+-\phi_1^-)\left(\Delta^+\left.{\partial \mathcal F \over \partial t}\right|_0\Delta^- \mathcal F \mbox{Re} Z_1 +
\Delta^-\left.{\partial \mathcal F \over \partial t}\right|_0\Delta^+ \mathcal F \mbox{Re} Z_1 \right. + \nonumber \\
&& \left.\left.\left.{\partial \mbox{Re} Z_1 \over \partial t}\right|_0\Delta^+ \mathcal F \Delta^- \mathcal F \right)\right) \nonumber
\end{eqnarray}
\begin{eqnarray}
\left. {\partial \mbox{Re} \varphi_1^1 \over \partial t}\right|_0 &=& {\eta_1\delta\over 2}\left[-\left((\phi_1^+)' \, \left.{\partial n^+ \over
\partial t}\right|_0 + (\phi_1^-)' \, \left.{\partial n^- \over \partial t}\right|_0\right)\Delta^- \mathcal F(\mbox{Im} M_1^+ (\mbox{Re} Z_1 - \vartheta_1\mbox{Im} Z_1)
+ \right.  \nonumber \\
&& \mbox{Re} M_1^+ (\mbox{Im} Z_1 + \vartheta_1 \mbox{Re} Z_1))  - (\phi_1^+ + \phi_1^-)\Delta^- \left.{\partial \mathcal F \over \partial
t}\right|_0 \times \nonumber \\
&& (\mbox{Im} M_1^+(\mbox{Re} Z_1 - \vartheta_1 \mbox{Im} Z_1) + \mbox{Re} M_1^+ (\mbox{Im} Z_1 + \vartheta_1 \mbox{Re} Z_1)) - \nonumber \\
&& (\phi_1^+ + \phi_1^-)\Delta^-\mathcal F\left(\left.{\partial \mbox{Im} M_1^+ \over \partial t}\right|_0(\mbox{Re} Z_1 - \vartheta_1 \mbox{Im} Z_1) \right. + \nonumber \\
&& \mbox{Im} M_1^+\left(\left.{\partial \mbox{Re} Z_1 \over \partial t}\right|_0-\left.{\partial \vartheta_1 \over \partial t}\right|_0\, \mbox{Im} Z_1 - \vartheta_1
\left.{\partial \mbox{Im} Z_1 \over \partial t}\right|_0\right) + \nonumber \\
&& \left.{\partial \mbox{Re} M_1^+ \over \partial t}\right|_0 (\mbox{Im} Z_1 + \vartheta_1 \mbox{Re} Z_1) + \mbox{Re} M_1^+\left(\left.{\partial \mbox{Im} Z_1 \over \partial
t}\right|_0 + \left.{\partial \vartheta_1 \over \partial t}\right|_0  \, \mbox{Re} Z_1 +  \right. \nonumber \\
&& \left. \left. \vartheta_1\left.{\partial \mbox{Re} Z_1 \over \partial t}\right|_0 \right)\right) + \left((\phi_1^+)'\left.{\partial n^+ \over
\partial t}\right|_0- (\phi_1^-)'\left.{\partial n^- \over \partial t}\right|_0\right)\Delta^+\mathcal F (\mbox{Re} Z_1-\vartheta_1 \mbox{Im} Z_1) + \nonumber \\
&& (\phi_1^+ - \phi_1^-)\left(\Delta^+\left.{\partial \mathcal F \over \partial t}\right|_0 (\mbox{Re} Z_1 - \vartheta_1 \mbox{Im} Z_1) + \right.\nonumber \\
&& \left.\left.\Delta^+ \mathcal F \left(\left.{\partial \mbox{Re} Z_1 \over \partial t}\right|_0 - \left.{\partial \vartheta_1 \over \partial t}\right|_0
\mbox{Im} Z_1 - \vartheta_1 \left.{\partial \mbox{Im} Z_1 \over \partial t}\right|_0\right)\right)\right] \nonumber \\
\left. {\partial \mbox{Im} \varphi_1^1 \over \partial t}\right|_0 &=& {\eta_1\delta\over 2}\left[\left((\phi_1^+)' \, \left.{\partial n^+ \over
\partial t}\right|_0 + (\phi_1^-)' \, \left.{\partial n^- \over \partial t}\right|_0\right)\Delta^- \mathcal F (\mbox{Re} M_1^+ (\mbox{Re} Z_1 - \vartheta_1 \mbox{Im} Z_1)
- \right.  \nonumber \\
&& \mbox{Im} M_1^+ (\mbox{Im} Z_1 + \vartheta_1 \mbox{Re} Z_1))  + (\phi_1^+ + \phi_1^-)\Delta^- \left.{\partial \mathcal F \over \partial t}\right|_0 \times \nonumber \\
&& (\mbox{Re} M_1^+(\mbox{Re} Z_1 - \vartheta_1 \mbox{Im} Z_1) - \mbox{Im} M_1^+ (\mbox{Im} Z_1 + \vartheta_1 \mbox{Re} Z_1)) - \nonumber \\
&& (\phi_1^+ + \phi_1^-)\Delta^-\mathcal F\left(\left.{\partial \mbox{Re} M_1^+ \over \partial t}\right|_0(\mbox{Re} Z_1 - \vartheta_1 \mbox{Im} Z_1) \right. + \nonumber \\
&& \mbox{Re} M_1^+\left(\left.{\partial \mbox{Re} Z_1 \over \partial t}\right|_0-\left.{\partial \vartheta_1 \over \partial t}\right|_0\, \mbox{Im} Z_1 - \vartheta_1
\left.{\partial \mbox{Im} Z_1 \over \partial t}\right|_0\right) - \nonumber \\
&& \left.{\partial \mbox{Im} M_1^+ \over \partial t}\right|_0 (\mbox{Im} Z_1 + \vartheta_1 \mbox{Re} Z_1) - \mbox{Im} M_1^+\left(\left.{\partial \mbox{Im} Z_1 \over \partial
t}\right|_0 + \left.{\partial \vartheta_1 \over \partial t}\right|_0  \, \mbox{Re} Z_1 +  \right. \nonumber \\
&& \left. \left. \vartheta_1\left.{\partial \mbox{Re} Z_1 \over \partial t}\right|_0 \right)\right) + \left((\phi_1^+)'\left.{\partial n^+ \over
\partial t}\right|_0- (\phi_1^-)'\left.{\partial n^- \over \partial t}\right|_0\right)\Delta^+\mathcal F (\vartheta_1 \mbox{Re} Z_1+ \mbox{Im} Z_1) + \nonumber \\
&& (\phi_1^+ - \phi_1^-)\left(\Delta^+\left.{\partial \mathcal F \over \partial t}\right|_0 (\vartheta_1 \mbox{Re} Z_1 +  \mbox{Im} Z_1) + \right.\nonumber \\
&& \left.\left.\Delta^+ \mathcal F \left(\vartheta_1\left.{\partial \mbox{Re} Z_1 \over \partial t}\right|_0 + \left.{\partial \vartheta_1 \over \partial
t}\right|_0 \mbox{Re} Z_1 +  \left.{\partial \mbox{Im} Z_1 \over \partial t}\right|_0\right)\right)\right] \nonumber
\end{eqnarray}
\begin{eqnarray}
\left. {\partial \mbox{Re} \varphi_1^2 \over \partial t}\right|_0 &=& {\eta_1\delta\over 2}\left[\left(\left((\phi_1^+)' \, \left.{\partial n^+ \over
\partial t}\right|_0 + (\phi_1^-)' \, \left.{\partial n^- \over \partial
t}\right|_0\right)\Delta^- \mathcal F + (\phi_1^++\phi_1^-)\Delta^- \left.{\partial \mathcal F \over \partial t}\right|_0 \right)\times \right. \nonumber \\
&&  (\mbox{Im} Z_1 + \vartheta_1 \mbox{Re} Z_1) + (\phi_1^++\phi_1^-)\Delta^- \mathcal F \left( \left.{\partial \mbox{Im} Z_1 \over \partial t}\right|_0 + \left.{\partial
\vartheta_1 \over \partial t}\right|_0 \mbox{Re} Z_1 + \vartheta_1 \left.{\partial \mbox{Re} Z_1 \over
\partial t}\right|_0 \right) +    \nonumber \\
&& \left(\left((\phi_1^+)' \, \left.{\partial n^+ \over \partial t}\right|_0 - (\phi_1^-)' \, \left.{\partial n^- \over
\partial t}\right|_0\right)\Delta^+ \mathcal F + (\phi_1^+-\phi_1^-)\Delta^+\left.{\partial \mathcal F \over \partial
t}\right|_0\right)\times    \nonumber \\
&& (\mbox{Re} M_1^-(\mbox{Re} Z_1- \vartheta_1 \mbox{Im} Z_1) - \mbox{Im} M_1^-(\mbox{Im} Z_1 + \vartheta_1 \mbox{Re} Z_1)) + \nonumber \\
&&(\phi_1^+-\phi_1^-)\Delta^+ \mathcal F \left(\left.{\partial \mbox{Re} M_1^- \over \partial t}\right|_0 (\mbox{Re} Z_1 -
\vartheta_1 \mbox{Im} Z_1) + \mbox{Re} M_1^-\left(\left.{\partial \mbox{Re} Z_1 \over \partial t}\right|_0 \right.\right. - \nonumber \\
&& \left. \left.{\partial \vartheta_1 \over \partial t}\right|_0 \mbox{Im} Z_1 - \vartheta_1 \left.{\partial \mbox{Im} Z_1 \over
\partial t}\right|_0 \right) - \left.{\partial \mbox{Im} M_1^- \over \partial t}\right|_0 (\mbox{Im} Z_1+\vartheta_1 \mbox{Re} Z_1) -  \nonumber \\
&&\left.\left. \mbox{Im} M_1^- \left( \left.{\partial \mbox{Im} Z_1 \over \partial t}\right|_0 + \left.{\partial \vartheta_1 \over
\partial t}\right|_0 \mbox{Re} Z_1 + \vartheta_1 \left.{\partial \mbox{Re} Z_1 \over \partial t}\right|_0 \right)\right)\right]
\nonumber\\
 \left. {\partial \mbox{Im} \varphi_1^2 \over \partial t}\right|_0 &=& {\eta_1\delta\over
2}\left[\left(\left((\phi_1^+)' \, \left.{\partial n^+ \over \partial t}\right|_0 + (\phi_1^-)' \, \left.{\partial n^- \over \partial
t}\right|_0\right)\Delta^- \mathcal F + (\phi_1^++\phi_1^-)\Delta^- \left.{\partial \mathcal F \over \partial t}\right|_0 \right)\times \right. \nonumber \\
&&  (-\mbox{Re} Z_1 + \vartheta_1 \mbox{Im} Z_1) + (\phi_1^++\phi_1^-)\Delta^- \mathcal F \left( -\left.{\partial \mbox{Re} Z_1 \over
\partial t}\right|_0 + \left.{\partial \vartheta_1 \over \partial t}\right|_0 \mbox{Im} Z_1 + \vartheta_1 \left.{\partial \mbox{Im}
Z_1 \over \partial t}\right|_0 \right) +    \nonumber \\
&& \left(\left((\phi_1^+)' \, \left.{\partial n^+ \over \partial t}\right|_0 - (\phi_1^-)' \, \left.{\partial n^- \over
\partial t}\right|_0\right)\Delta^+ \mathcal F + (\phi_1^+-\phi_1^-)\Delta^+\left.{\partial \mathcal F \over \partial
t}\right|_0\right)\times    \nonumber \\
&& (\mbox{Re} M_1^-(\mbox{Im} Z_1+ \vartheta_1 \mbox{Re} Z_1) + \mbox{Im} M_1^-(\mbox{Re} Z_1 - \vartheta_1 \mbox{Im} Z_1)) + \nonumber \\
&&(\phi_1^+-\phi_1^-)\Delta^+ \mathcal F \left(\left.{\partial \mbox{Re} M_1^- \over \partial t}\right|_0 (\mbox{Im} Z_1 +
\vartheta_1 \mbox{Re} Z_1) + \mbox{Re} M_1^-\left(\left.{\partial \mbox{Im} Z_1 \over \partial t}\right|_0 \right.\right. + \nonumber \\
&& \left. \left.{\partial \vartheta_1 \over \partial t}\right|_0 \mbox{Re} Z_1 + \vartheta_1 \left.{\partial \mbox{Re} Z_1 \over
\partial t}\right|_0 \right) + \left.{\partial \mbox{Im} M_1^- \over \partial t}\right|_0 (\mbox{Re} Z_1-\vartheta_1 \mbox{Im} Z_1) +  \nonumber \\
&&\left.\left. \mbox{Im} M_1^- \left( \left.{\partial \mbox{Re} Z_1 \over \partial t}\right|_0 - \left.{\partial \vartheta_1 \over
\partial t}\right|_0 \mbox{Im} Z_1 - \vartheta_1 \left.{\partial \mbox{Im} Z_1 \over \partial t}\right|_0 \right)\right)\right] \nonumber
\end{eqnarray}
\begin{eqnarray}
\left. {\partial \mbox{Re} \varphi_1^3 \over \partial t}\right|_0 &=& {1\over 2}\left((\phi_1^+)' \, \left.{\partial n^+
\over \partial t}\right|_0  -  (\phi_1^-)' \, \left.{\partial n^- \over \partial t}\right|_0  \right)\times \nonumber \\
&& \left[{1 \over 1+\vartheta_1^2} - \eta_1\delta^2(\Delta^+ \mathcal F)^2\,(\mbox{Re} M_1^- \mbox{Re} Z_1 - \mbox{Im} M_1^- \mbox{Im} Z_1)\right] +\nonumber \\
&& {1\over 2}(\phi_1^+ - \phi_1^-)\left[{-2\vartheta_1 \over (1+\vartheta_1^2)^2} \left.{\partial \vartheta_1 \over
\partial t}\right|_0 \right. - \nonumber \\
&&   2\eta_1\delta^2 \Delta^+ \mathcal F \Delta^+ \left.{\partial \mathcal F \over \partial t}\right|_0 (\mbox{Re} M_1^- \mbox{Re}
Z_1 - \mbox{Im} M_1^- \mbox{Im} Z_1) -  \nonumber \\
&&  \eta_1\delta^2 (\Delta^+ \mathcal F)^2\left(\left.{\partial \mbox{Re} M_1^- \over \partial t}\right|_0 \mbox{Re} Z_1 - \mbox{Re} M_1^-
\left.{\partial \mbox{Re} Z_1 \over \partial t}\right|_0 - \mbox{Im} Z_1 \left.{\partial \mbox{Im} M_1^- \over \partial t}\right|_0 \right. - \nonumber \\
&& \left.\left. \mbox{Im} M_1^- \left.{\partial \mbox{Im} Z_1 \over \partial t}\right|_0\right) \right] - \nonumber \\
&& {\eta_1\delta^2 \over 2}\left(\left((\phi_1^+)' \left.{\partial n^+ \over \partial t}\right|_0 + (\phi_1^-)'
\left.{\partial n^- \over \partial t}\right|_0 \right) \Delta^+ \mathcal F \Delta^- \mathcal F \mbox{Im} Z_1 + \right.\nonumber \\
&&(\phi_1^++\phi_1^-)\left(\Delta^+\left.{\partial \mathcal F \over \partial t}\right|_0\Delta^- \mathcal F \mbox{Im} Z_1 +
\Delta^-\left.{\partial \mathcal F \over \partial t}\right|_0\Delta^+ \mathcal F \mbox{Im} Z_1 \right. + \nonumber \\
&& \left.\left.\left. {\partial \mbox{Im} Z_1 \over \partial t}\right|_0\Delta^+ \mathcal F \Delta^- \mathcal F \right)\right) \nonumber \\
\left. {\partial \mbox{Im} \varphi_1^3 \over \partial t}\right|_0 &=& {1\over 2}\left((\phi_1^+)' \, \left.{\partial n^+
\over \partial t}\right|_0  +  (\phi_1^-)' \, \left.{\partial n^- \over \partial t}\right|_0  \right)\times \nonumber \\
&& \left[{-\vartheta_1 \over 1+\vartheta_1^2} - \eta_1\delta^2(\Delta^+ \mathcal F)^2\,(\mbox{Re} M_1^- \mbox{Im} Z_1 + \mbox{Im} M_1^- \mbox{Re} Z_1)\right] +\nonumber \\
&& {1\over 2}(\phi_1^+ - \phi_1^-)\left[{\vartheta_1^2-1 \over (1+\vartheta_1^2)^2} \left.{\partial \vartheta_1 \over
\partial t}\right|_0 \right. - \nonumber \\
&&   2\eta_1\delta^2 \Delta^+ \mathcal F \Delta^+ \left.{\partial \mathcal F \over \partial t}\right|_0 (\mbox{Re} M_1^- \mbox{Im}
Z_1 + \mbox{Im} M_1^- \mbox{Re} Z_1) -  \nonumber \\
&&  \eta_1\delta^2 (\Delta^+ \mathcal F)^2\left(\left.{\partial \mbox{Re} M_1^- \over \partial t}\right|_0 \mbox{Im} Z_1 + \mbox{Re} M_1^-
\left.{\partial \mbox{Im} Z_1 \over \partial t}\right|_0 + \mbox{Re} Z_1 \left.{\partial \mbox{Im} M_1^- \over \partial t}\right|_0 \right. + \nonumber \\
&& \left.\left. \mbox{Im} M_1^- \left.{\partial \mbox{Re} Z_1 \over \partial t}\right|_0\right) \right] + \nonumber \\
&& {\eta_1\delta^2 \over 2}\left(\left((\phi_1^+)' \left.{\partial n^+ \over \partial t}\right|_0 + (\phi_1^-)'
\left.{\partial n^- \over \partial t}\right|_0 \right) \Delta^+ \mathcal F \Delta^- \mathcal F \mbox{Re} Z_1 + \right.\nonumber \\
&& (\phi_1^++\phi_1^-)\left(\Delta^+\left.{\partial \mathcal F \over \partial t}\right|_0\Delta^- \mathcal F \mbox{Re} Z_1 +
\Delta^-\left.{\partial \mathcal F \over \partial t}\right|_0\Delta^+ \mathcal F \mbox{Re} Z_1 \right. + \nonumber \\
&& \left.\left.\left.{\partial \mbox{Re} Z_1 \over \partial t}\right|_0\Delta^+ \mathcal F \Delta^- \mathcal F \right)\right) \nonumber
\end{eqnarray}
In the above expressions we have used $(\phi_1^\pm)'= \partial \phi_1^\pm / \partial n^\pm $ and $\left. \partial \vartheta_1 /
\partial t \right|_0 = \left<\left. \partial \mathcal F / \partial t\right|_0\right>_1$. We also need to calculate $Z_j$, $M_j^\pm$,
$\left.{\partial Z_1 / \partial t}\right|_0$ and $\left.{\partial M_1^\pm /\partial t}\right|_0$:
\begin{eqnarray}
\mbox{Re} Z_j &=& {Z_{j1} \over Z_{j1}^2 + Z_{j2}^2} \nonumber \\
\mbox{Im} Z_j &=& {Z_{j2} \over Z_{j1}^2 + Z_{j2}^2} \nonumber
\end{eqnarray}
Where the functionals $Z_{j1}(F)$ and $Z_{j2}(F)$ are as follows:
\begin{eqnarray}
 Z_{j1} &=& 1 - 6\vartheta_j^2 +\vartheta_j^4 + (1-\vartheta_j^2)\left(\eta_1^2 + \delta^2\left((\Delta_j^- \mathcal F)^2 + (\Delta_j^+ \mathcal
F)^2\right)\right)+ \delta^4\left(\Delta_j^- \mathcal F \, \Delta_j^+ \mathcal F\right)^2 \nonumber \\
Z_{j2} &=& -2\vartheta_j\left(2 + \eta_1^2 + \delta^2\left((\Delta_j^- \mathcal F)^2 + (\Delta_j^+ \mathcal F)^2\right) - 2\vartheta_j^2\right)
\nonumber
\end{eqnarray}
Therefore, the time derivative $\displaystyle\left.{\partial Z_{1} \over \partial t}\right|_0$ is as follows:
\begin{eqnarray}
 \left.{\partial \mbox{Re} Z_{1} \over \partial t}\right|_0 &=& { (Z_{12}^2-Z_{11}^2)\displaystyle\left.{\partial Z_{11} \over \partial t}\right|_0 -
 2Z_{11}Z_{12}\left.{\partial Z_{12} \over \partial t}\right|_0 \over (Z_{11}^2 + Z_{12}^2)^2} \nonumber \\
\left.{\partial \mbox{Im} Z_{1} \over \partial t}\right|_0 &=& { (Z_{11}^2-Z_{12}^2)\displaystyle\left.{\partial Z_{12} \over \partial t}\right|_0 -
 2Z_{11}Z_{12}\left.{\partial Z_{11} \over \partial t}\right|_0 \over (Z_{11}^2 + Z_{12}^2)^2} \nonumber
\end{eqnarray}
Where the time derivatives $\displaystyle\left.{\partial Z_{11} \over \partial t}\right|_0$ and $\displaystyle\left.{\partial Z_{12} \over \partial
t}\right|_0$ are as follows:
\begin{eqnarray}
 \left.{\partial Z_{11} \over \partial t}\right|_0 &=& 4( \vartheta_1^3 - 3 \vartheta_1)\left.{\partial \vartheta_{1} \over \partial t}\right|_0 +
 2(1- \vartheta_1^2)\left(\delta^2 \left(\Delta^+ \mathcal F \Delta^+ \left.{\partial \mathcal F \over \partial t}\right|_0 + \Delta^-  \mathcal F
 \Delta^- \left.{\partial \mathcal F \over \partial t}\right|_0\right) - \right. \nonumber \\
 &&\left. \vartheta_1 \left.{\partial \vartheta_1 \over \partial t}\right|_0(\eta_1^2 + \delta^2 ((\Delta^- \mathcal F)^2 +
 (\Delta^+ \mathcal F)^2)\right) + \nonumber \\
 && 2\delta^4 \Delta^+ \mathcal F \Delta^- \mathcal F \left(\Delta^+ \mathcal F\, \Delta^- \left.{\partial \mathcal F \over \partial t}\right|_0 +
  \Delta^- \mathcal F\, \Delta^+ \left.{\partial \mathcal F \over \partial t}\right|_0\right) \nonumber \\
 \left.{\partial Z_{12} \over \partial t}\right|_0 &=& -2\left[\left.{\partial \vartheta_{1} \over \partial t}\right|_0
 (2 + \eta_1^2 + \delta^2\left((\Delta^- \mathcal F)^2 + (\Delta^+ \mathcal F)^2\right) - 2\vartheta_1^2) \, +  \right. \nonumber \\
 && \left. 2\vartheta_1 \left(\delta^2\left(\Delta^+ \mathcal F\, \Delta^- \left.{\partial \mathcal F \over \partial t}\right|_0 +
  \Delta^- \mathcal F\, \Delta^+ \left.{\partial \mathcal F \over \partial t}\right|_0\right) -
  2\vartheta_1  \left.{\partial \vartheta_1 \over \partial t}\right|_0\right)\right] \nonumber
\end{eqnarray}
The functionals $M_j^\pm(F)$ are as follows:
\begin{eqnarray}
\mbox{Re} M_j^\pm &=& {1 \over \eta_1}\left[1 + {\delta^2 (\Delta_j^\pm \mathcal F)^2 \over 1+ \vartheta_j^2} \right] \nonumber \\
\mbox{Im} M_j^\pm &=& {1 \over \eta_1}\left[\vartheta_j - {\delta^2 \vartheta_j (\Delta_j^\pm \mathcal F)^2 \over 1+ \vartheta_j^2} \right] \nonumber
\end{eqnarray}
Finally, we need to calculate the time derivative $\displaystyle\left.{\partial M_1^\pm \over \partial t}\right|_0$:
\begin{eqnarray}
 \left.{\partial \mbox{Re} M_1^\pm \over \partial t}\right|_0 &=& {2\delta^2 \Delta^\pm \mathcal F\over \eta_1 (1+\vartheta_1^2)}\left( \Delta^\pm
 \left.{\partial \mathcal F \over \partial t}\right|_0 - {\vartheta_1 \Delta^\pm \mathcal F \over 1+\vartheta_1^2}\left.{\partial \vartheta_1 \over
 \partial t}\right|_0 \right) \nonumber \\
 \left.{\partial \mbox{Im} M_1^\pm \over \partial t}\right|_0 &=& {1 \over \eta_1}\left[ \left.{\partial \vartheta_1 \over
 \partial t}\right|_0 - {\delta^2 \Delta^\pm \mathcal F \over (1+\vartheta_1^2)^2} \left( (1-\vartheta_1^2) \Delta^\pm \mathcal F  \left.{\partial \vartheta_1 \over
 \partial t}\right|_0 + 2\vartheta_1 (1+\vartheta_1^2) \Delta^\pm  \left.{\partial \mathcal F \over \partial t}\right|_0 \right) \right] \nonumber
\end{eqnarray}

\end{document}